\documentclass[manuscript, screen]{acmart}

\AtBeginDocument{%
	\providecommand\BibTeX{{%
			\normalfont B\kern-0.5em{\scshape i\kern-0.25em b}\kern-0.8em\TeX}}}

\usepackage{fancyhdr}
\usepackage[ruled, linesnumbered]{algorithm2e}
\usepackage{graphicx}
\usepackage{subfigure}
\usepackage{url}
\usepackage{multirow}
\usepackage{array}
\usepackage{booktabs}
\usepackage{enumitem}
\usepackage{rotating}

\usepackage{amsmath,amssymb,amsthm}  

\usepackage{xcolor}
\usepackage{color}

\usepackage{listings}
\usepackage{caption}
\usepackage{tcolorbox}

\setcopyright{acmcopyright}
\acmJournal{TOSEM}

\setcopyright{none}
\renewcommand\footnotetextcopyrightpermission[1]{} 
\begin{document}

\title{Comparing One with Many --- Solving Binary2source Function Matching Under Function Inlining}

\author{Ang Jia}
\email{jiaang@stu.xjtu.edu.cn}
\author{Ming Fan}
\email{mingfan@mail.xjtu.edu.cn}
\author{Xi Xu}
\email{xx19960325@stu.xjtu.edu.cn}
\author{Wuxia Jin}
\email{jinwuxia@mail.xjtu.edu.cn}
\author{Haijun Wang}
\email{hjwang.china@gmail.com}

\affiliation{%
	\institution{Xi'an Jiaotong University}
	\state{Shaanxi}
	\city{Xi'an}
	\country{China}
	\postcode{710049}
}

\author{Qiyi Tang}
\email{dodgetang@tencent.com}
\author{Sen Nie}
\email{snie@tencent.com}
\author{Shi Wu}
\email{shiwu@tencent.com}
\affiliation{%
	\institution{Tencent Security Keen Lab}
	\city{Shanghai}
	\country{China}
	\postcode{710049}
}

\author{Ting Liu}
\email{tingliu@mail.xjtu.edu.cn}
\affiliation{%
	\institution{Xi'an Jiaotong University}
	\city{Xi'an}
	\country{China}
}

\renewcommand{\shortauthors}{Ang Jia et al.}

\begin{abstract}
Binary2source function matching is a fundamental task for many security applications, including Software Component Analysis (SCA). \textbf{``1-to-1''} mechanism has been applied in existing binary2source matching works, in which one binary function is matched against one source function. However, we discovered that such mapping could be \textbf{``1-to-n''} (one query binary function maps multiple source functions), due to the existence of \textbf{function inlining}.

To help conduct binary2source function matching under function inlining, we propose a method named \textbf{O2NMatcher} to generate Source Function Sets (SFSs) as the matching target for binary functions with inlining. 
We first propose a model named \textbf{ECOCCJ48} for inlined call site prediction. To train this model, we leverage the compilable OSS to generate a dataset with labeled call sites (inlined or not), extract several features from the call sites, and design a compiler-opt-based multi-label classifier by inspecting the inlining correlations between different compilations. Then, we use this model to predict the labels of call sites in the uncompilable OSS projects without compilation and obtain the labeled function call graphs of these projects. Next, we regard the construction of SFSs as a sub-tree generation problem and design root node selection and edge extension rules to construct SFSs automatically. Finally, these SFSs will be added to the corpus of source functions and compared with binary functions with inlining. We conduct several experiments to evaluate the effectiveness of O2NMatcher and results show our method increases the performance of existing works by 6\% and exceeds all the state-of-the-art 
works.\footnote{New Paper}
\end{abstract}



\maketitle

\section{Introduction}
\label{sec:introduction}


Most software today is not developed entirely from scratch. Instead, developers rely on a range of open-source components to create their applications\cite{kula2018developers}. According to a report published by Gartner \cite{Gartner_SCA}, over 90\% of the development organizations stated that they rely on open-source components. Although using open-source components helps to finish projects quicker and reduce costs, dependence on risky open-source components brings software supply chain security risks\cite{gkortzis2021software}. For example, due to code reuse, a single vulnerability (e.g., the Heartbleed \cite{heartbleed} in OpenSSL \cite{openssl}) may spread across thousands of software, causing 17\% (around half a million) of the Internet's secure web servers vulnerable.

To avoid vulnerable dependence, Software Component Analysis (SCA) \cite{chen2020automated} is proposed to discover software's dependence on Open Source Software (OSS) projects. Usually, the SCA service provider maintains a large OSS codebase. When commercial software companies send their released binary executables, the SCA service provider compares these binaries with the OSS projects and returns a report of the OSS components that the queried executables contain.


The comparison procedure in the SCA is accomplished by binary2source function matching. When a query binary is provided, this query binary will be first dissembled into binary functions and then these binary functions will be compared with the source functions of the OSS. Existing binary2source matching works, such as B2SMatcher \cite{ban2021b2smatcher} and CodeCMR \cite{codecmr}, conduct an \textbf{``1-to-1''} matching between the query binary function and the source function. However, we recover that such mapping could be \textbf{``1-to-n''} (one query binary function maps multiple source functions), due to the existence of \textbf{function inlining}.

\begin{figure}[htbp]
	\centering
	\includegraphics[width=0.8\textwidth]{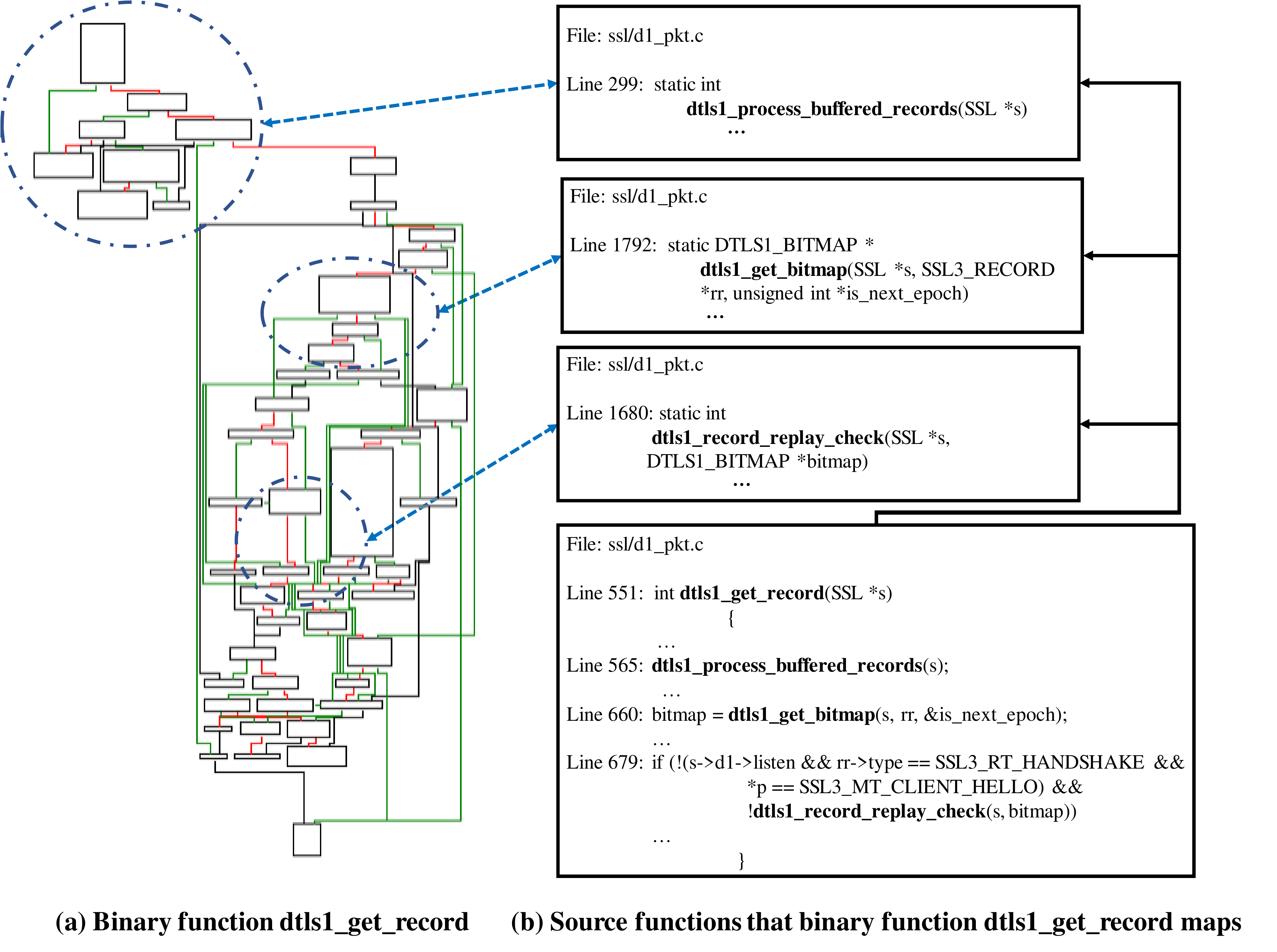}
	\vspace{-8pt}
	\caption{``1-to-n'' matching cases brought by function inlining}
	\label{fig:inlining_example}
	\vspace{-10pt}
\end{figure}

Figure \ref{fig:inlining_example} shows an ``1-to-n'' matching case when comparing the binary function \textit{dtls1\_get\_record} with the source function \textit{dtls1\_get\_record}. Source function \textit{dtls1\_get\_record} is from OpenSSL in version 1.0.1j which is associated with the vulnerability CVE-2014-3571\cite{CVE_example} and binary function \textit{dtls1\_get\_record} is compiled from it using gcc-8.2.0 with O2. 
But when we use CodeCMR to search the corresponding source function for binary function \textit{dtls1\_get\_record}, we find that the similarity between the binary function \textit{dtls1\_get\_record} and the source function \textit{dtls1\_get\_record} is 
less than 60\%, leading a failure in fetching the corresponding source function in the top 10 returns. 
This mismatch causes existing works to fail to detect this vulnerability and leaves this binary function risky.


This is the issue that \textbf{function inlining} has brought. When we look deeper into the compilation, we find that source function \textit{dtls1\_get\_record} inlines the contents of several other source functions, including \textit{dtls1\_process\_buffered\_records}, \textit{dtls1\_get\_bitmap} and \textit{dtls1\_record\_replay\_check}, into its body. Function inlining, which replaces a function call with the embedded function body, greatly changes the content of the function, thus affecting the progress of function matching. As a result, source function \textit{dtls1\_get\_record} and binary function \textit{dtls1\_get\_record} become less similar, and existing ``1-to-1'' matching mechanism suffers when  conducting binary2source function matching under function inlining.


As function inlining widely exists in binaries \cite{jia2021one}, It is important to resolve the issues function inlining brings. However, currently, there are still three challenges for conducting binary2source matching under function inlining.

\textbf{C1: ``Out-of-domain'' (OOD) issue in OSS projects.} 
The binary functions with inlining are generated by more than one source function. However, there does not exist a source function in the OSS projects that has the fully same semantics as the queried binary function. As a result, the query of binary function with inlining is experiencing an OOD issue, where an exact match of the queried binary function does not exist in the OSS corpus.

\textbf{C2: Opaque inlining in stripped binaries.} The queried binaries are often stripped which makes it harder to infer which binary functions are generated with inlining and which source functions are inlined into the binary function. The opaque inlining makes it hard to decide when and how to apply a strategy to tackle function inlining.

\textbf{C3: Various inlining decisions in different binaries.}
The queried binaries are generated through enormous compilation settings and thus result in various inlining decisions in the release binaries. It is difficult to cover all the inlining cases without a  comprehensive study of the correlation between inlining under various compilation settings.



\begin{figure}[h]
	\centering
	\vspace{-5pt}
	\includegraphics[width=0.65\textwidth]{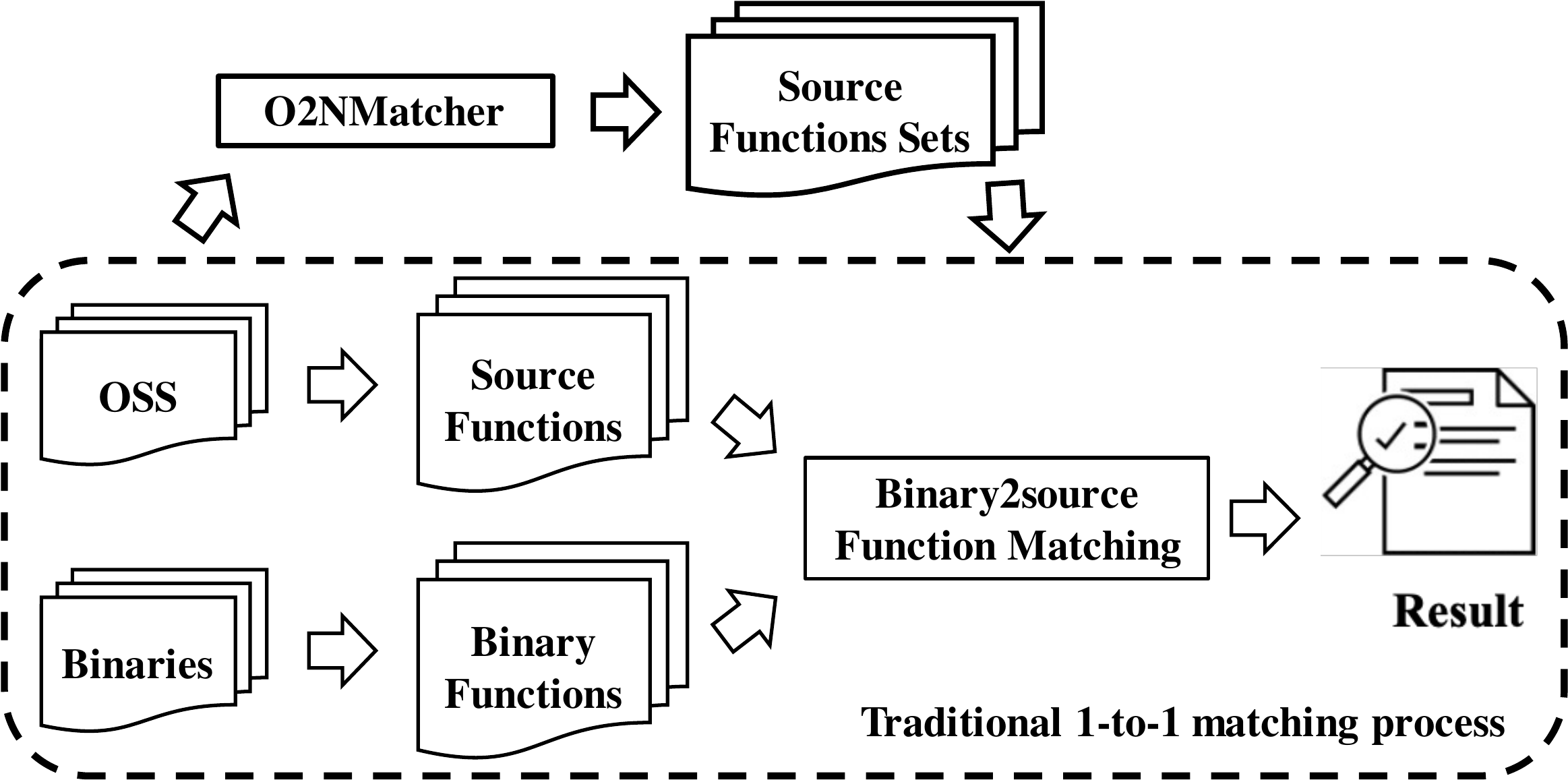}
	\vspace{-5pt}
	\caption{The key idea of O2NMatcher}
	\label{fig:overview_old}
	\vspace{-5pt}
\end{figure}

To tackle the aforementioned three challenges, we propose a matching method named \textbf{O2NMatcher} that uses the \textbf{``1-to-n''} \textbf{matching} mechanism to match binary functions with source functions. 
Instead of developing an end-to-end method for binary2source matching, our method works on the input source functions and aims to help existing works to promote their effectiveness under function inlining. 
Figure \ref{fig:overview_old} shows the key idea of O2NMatcher. Generally, existing works directly compare binary functions and source functions to obtain their similarities. However, this ``1-to-1'' matching mechanism misses the semantics of the source functions that are inlined into the binary functions. Our method aims to generate SFSs as a supplement to the source functions. Comparing source functions with binary functions solves the ``1-to-1'' matching tasks, while comparing SFSs with binary functions solves the ``1-to-n'' matching tasks.

To tackle \textbf{C1}, we focus on generating Source Function Sets (SFSs) to complement the corpus for queries of binary functions with inlining. For example, the source functions
\textit{dtls1\_get\_record}, \textit{dtls1\_process\_buffered\_records}, \textit{dtls1\_get\_bitmap} and \textit{dtls1\_record\_replay\_check} together compose the SFS for the queried binary function \textit{dtls1\_get\_record} shown in Figure \ref{fig:inlining_example}. 
To tackle \textbf{C2}, a classifier for Inlined Call Site (ICS) prediction is trained on the compilable OSS projects and is used to predict ICSs in the uncompilable OSS projects. To support the training process, we propose a dataset labeling method to automatically identify the inlined call sites in the binaries.
To tackle \textbf{C3}, the classifier is trained towards enormous compilation setting combinations. We model it as a Multi-Label Classification (MLC) problem and propose a compiler-opt-based classifier based on the correlations of inlining decisions under different compilation settings. 



Through our method, SFSs are generated for the given OSS without compilation and then these SFSs are compared with binary functions using existing ``1-to-1'' methods.
We conduct several experiments to evaluate the effectiveness of our method. 
Results show our method can achieve a 6\% improvement in recall when detecting the unseen OSS and exceeds all state-of-the-art methods.

The major contributions are as follows:

\begin{itemize}
	\item To the best of our knowledge, we are the first to investigate the solution for binary2source matching under function inlining.
	
	\item We conduct several studies to analyze the inlining correlations between different compilation settings, which help us model the prediction of inlined call sites as a multi-label classification problem and propose a compiler-opt-based classifier named ECOCCJ48.
	
	\item Based on ECOCCJ48, we propose a method named O2NMatcher that uses ``1-to-n'' matching to tackle function inlining. Our generation method can automatically generate SFSs for binary functions with inlining.
	
	\item We evaluate O2NMatcher through several experiments and results show our method can achieve a 6\% improvement in detecting inlined functions for existing binary2source matching works.

\end{itemize}

To facilitate further research, we have made the source code and dataset publicly available \cite{dataset}.

\section{Overview}
\label{sec:overview}

In this section, we will introduce the workflow of O2NMatcher.

\begin{figure*}[h]
	\centering
	\vspace{-5pt}
	\includegraphics[width=0.85\textwidth]{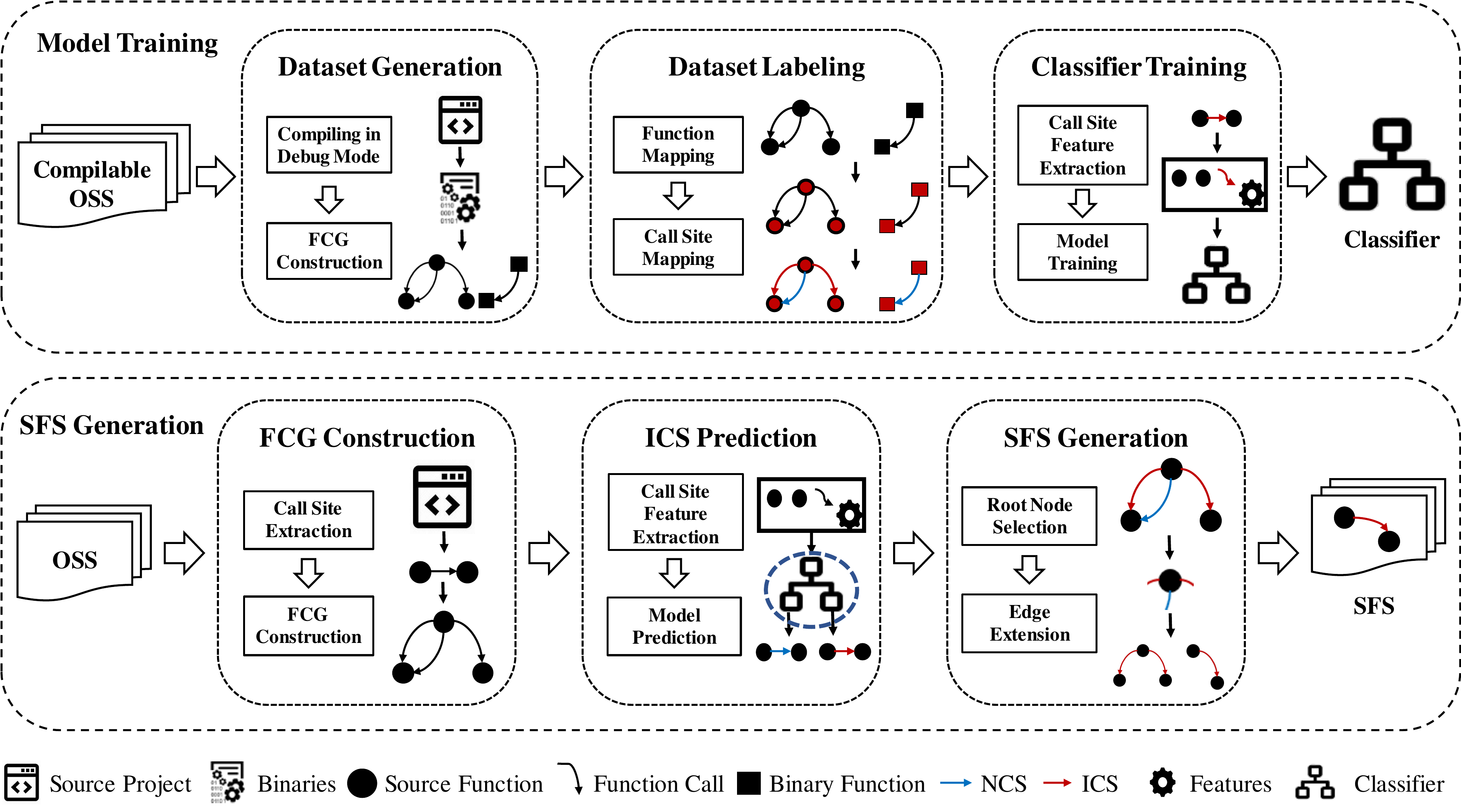}
	\vspace{-8pt}
	\caption{Workflow of O2NMatcher}
	\label{fig:overview}
	\vspace{-7pt}
\end{figure*}

Figure \ref{fig:overview} shows the overall workflow of O2NMatcher. The workflow of O2NMatcher is divided into two parts: Model Training and SFS Generation. 

The target of model training is to generate a classifier to predict ICSs in uncompilable OSS projects. The model is trained on the compilable OSS on the source level and can be applied to uncompilable OSS projects without compilation. 
The model training process is divided into three steps: Dataset Generation, Dataset Labeling, and Classifier Training. Firstly, a dataset is generated by compiling OSS projects to binaries enabled with function inlining. Then, the inlined call sites in the binaries are labeled to serve as the training set for the classifier. Finally, a classifier is designed and trained on the labeled dataset, which produces the classifier used in the SFS generation process.

The SFS generation is based on the trained model and leverages it to generate SFSs for uncompilable OSS projects. The SFS generation process is divided into three steps: FCG (Function Call Graph) Construction, ICS Prediction, and SFS Generation. Firstly, the FCG Construction process identifies call sites in the OSS and constructs FCGs for projects. Then, the ICS Prediction process extracts features for all call sites and uses the trained model to predict which call sites will be inlined during compilation. Finally, the SFS Generation process generates the SFSs based on the inlined call sites. 




The following two sections will illustrate the full progress of our O2NMatcher. In detail, Section \ref{sec:classifier} presents the process of Model Training and Section \ref{sec:SFS_generation} presents the process of SFS Generation.


\section{Model Training process of O2NMatcher}
\label{sec:classifier}

\subsection{Dataset Generation and Labeling}

Dataset generation and labeling are the bases for training a model. Generally, we propose a method, which leverages the debug information generated by Dwarf \cite{Dwarf_v5} during the compilation, to label the dataset. It will generate the binary2source line-level and function-level mappings together with the labels (inlined or not) of the call sites.

In detail, when compiling the source projects with the ``-g'' option, a \textit{.debug\_line} section, which contains the mapping from the binary address to the source file line (line-level mapping), will be produced. Similar to the idea in \cite{jia2021one}, we extend the line-level mapping to the function-level mapping by using existing tools, such as Understand \cite{understand} and IDA Pro \cite{IDAPro} or tree-sitter \cite{tree-sitter} and Ghidra \cite{Ghidra}. 

However, though function-level mapping directly identifies inlined source functions (function mapping that a binary function is mapped to multiple source functions will be identified as an inlining case), it is not suitable for training a classifier that directly predicts which functions will be inlined. That is because the compilers decide whether to inlining on the features of each call site. Callees that are inlined into one of their callers may not be inlined into another. Thus, for the accuracy of ICS prediction,  our labeling process further identifies which call site is inlined.

\begin{figure}[ht]
	\centering
	\vspace{-3pt}
	\subfigure[Easily inferred call sites]{
		\begin{minipage}[t]{0.35\linewidth}
			\centering
			\includegraphics[width=0.9\textwidth]{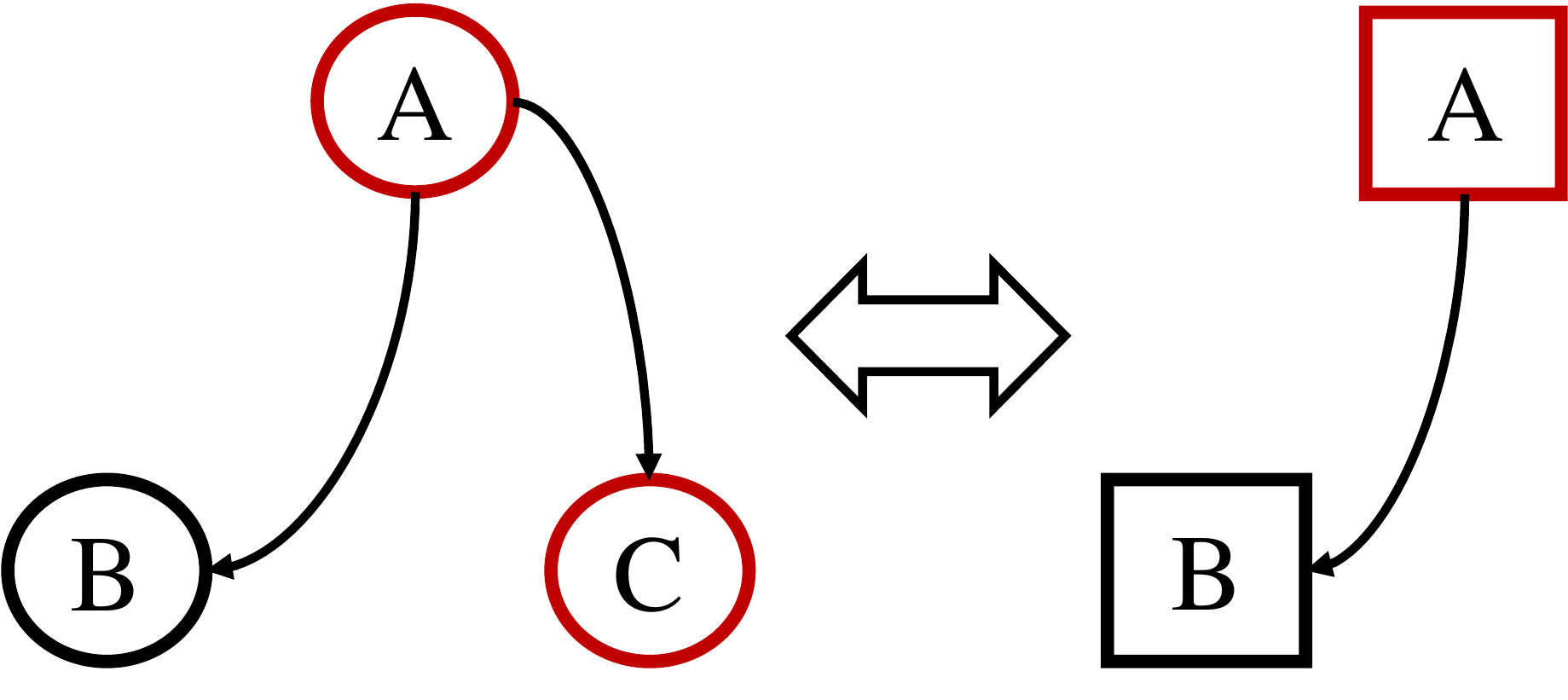}
			\label{fig:labeling_1}
			\vspace{-10pt}
		\end{minipage}%
	}%
	\centering
	\subfigure[Confused call sites]{
		\begin{minipage}[t]{0.35\linewidth}
			\centering
			\includegraphics[width=0.9\textwidth]{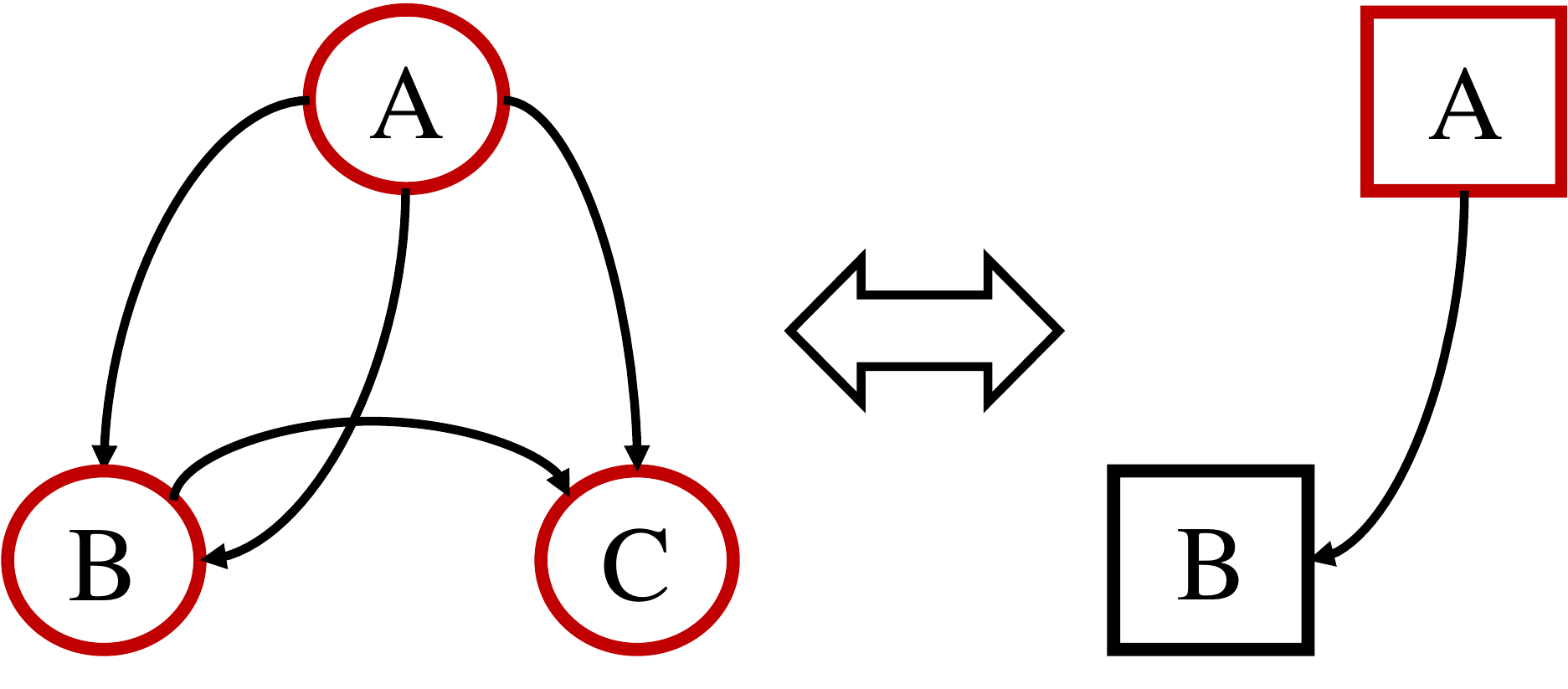}
			\label{fig:labeling_2}
			\vspace{-10pt}
		\end{minipage}%
	}%
	\vspace{-12pt}
	\caption{Inferring inlined call sites from function-level mapping}
	\label{fig:labeling}
	\vspace{-7pt}
\end{figure}

Labels of some call sites sometimes can be inferred directly by their function mapping. Figure \ref{fig:labeling} shows two examples shown in FCG. In the source and binary FCGs, the circle represents the source function while the square represents the binary function. A red square and several red circles represent the binary function and its mapped source function. In Figure \ref{fig:labeling_1}, binary function A is mapped to source functions A and C. As A only has one call to C, thus the call site from A to C is inlined and other call sites are not. However, Figure \ref{fig:labeling_2} presents an example that may have multiple inlining decisions. For example, as A calls B twice, it is hard to infer which call site is inlined. Moreover, A calls C and B calls C, it is also difficult to decide which call site inlines the function C.  

To identify the inlined call sites, we further analyze the line mapping of these call sites. Simply, when there are binary call sites whose address is mapped to the source call site, this source call site is not inlined. Otherwise, this source call site is inlined. 

As our labeling method only depends on the debug information, our labeling method can be easily applied to any compiled projects that are compiled in debug mode. Considering that the effectiveness of the inlining identification method has also been evaluated and discussed in \cite{jia2021one}, we will not repeat it here.

\subsection{Call Site Feature Extraction}
\label{sec:CS_feature}
After obtaining the labels for call sites, it is important to select appropriate features for these call sites as the input of the classifier. Considering that compilers usually decide whether to inline by measuring its cost and benefit, we summarize several features that can measure the cost and benefit of inlining a call site.

\begin{table}[h]
	\vspace{-5pt}
	\caption{Selected features for ICS prediction}
	\vspace{-8pt}
	\centering
	\scalebox{0.9}{
		\begin{tabular}{c|c|c}
            \hline
            Type                              & Part                & \multicolumn{1}{c}{Feature}                                                    \\ \hline
            \multirow{3}{*}{Caller/Callee}    & Function body       & Statement, While, Switch, Switch Cases, If, For, Return, Declare, Expression    \\ \cline{2-3} 
                                              & Function definition & Inline keyword, Static keyword                                                  \\ \cline{2-3} 
                                              & Function call       & Calling times, Called times                                                     \\ \hline
            \multirow{2}{*}{Call Instruction} & Location            & Path length, Whether in for, Whether in while, Whether in switch, Whether in if \\ \cline{2-3} 
                                              & Arguments           & Number of arguments, Number of constant arguments                               \\ \hline
        \end{tabular}}
	\label{tab:features}
	\vspace{-5pt}
\end{table}

Table \ref{tab:features} shows the selected features for call sites. Generally, the features mainly come from two parts: Caller/Callee and Call Instruction. 

\subsubsection{Caller and Callee} 
For caller and callee, we extract features from their function bodies, their function definition, and their function calls. These features reflect the possible cost and benefit of inlining a callee function.

Firstly, in the function body, we use a source parser to parse instructions in the function and count different kinds of instructions. As shown in Table \ref{tab:features}, \textit{statement} means the number of all statements and \textit{while} means the number of while statements. The count of these instructions represents the size of a function, which indicates the cost of inlining a callee. Simply, inlining a small callee function only brings a little size increase to the caller function while a large callee function may raise a high size-increasing cost which prevents the inlining.

Then, in the function definition, we focus on two important keywords, \textit{inline} and \textit{static}. Keyword \textit{inline} is a suggestion to the compiler that when other functions call this function, this function should be inlined. Functions decorated by \textit{static} are only accessed in the same translation unit. These keywords either directly influence the inlining decisions of the compiler, or indirectly influence the FCG around the functions, which are important features for ICS prediction.

Finally, we extract the calling times and called times of the caller and callee which may affect the inlining benefit. Intuitively, if a callee function is called by enormous caller functions, the cost of inlining it into all its callers will increase with the number of caller functions. However, when a function is called only once, inlining will only bring the benefit of further optimization without the size increase, which is an ideal case to conduct inlining. Thus, we extract their calling and called times to facilitate further prediction.

\subsubsection{Call Instruction} 
For the call sites, we focus on their locations and arguments which affects the inlining benefit. For example, if a call site is in a loop defined by \textit{for} or \textit{while}, inlining this call site will greatly reduce the time of function call. To record the situation that a call site may exist in nest loops, we record the \textit{path length} from the function entry to the call site. Besides, if a call site contains a constant argument, this argument may help reduce the inlined function size as this argument may help determine some branches in the callee functions. Thus, features from the arguments are also extracted to facilitate the ICS prediction.


\subsection{Model Training}

Based on the labeled dataset, we can train a model to learn how the compiler decides which call sites will be inlined. However, this is not simple as replicating the strategies used in the compilers, considering the following reasons:

\begin{itemize}
	\item [(1)]
	Compilers usually make inlining decisions at the intermediate language and consider the overall inlining gain and cost, while our model can only parse the source code considering that our model is designed for uncompilable source projects. The same estimation used in compilers cannot be directly applied to source code.
	
	\item [(2)]
	Different compilers and different optimizations will bring out different inlining decisions. Copying the strategies used in the compiler is labor-intensive and cannot be extended to new compilers.
	
	\item [(3)]
	The inlining decisions made by the compilers under different optimizations are usually associated with each other. For example, GCC applies most optimizations in O2 when using the O3 option. Capturing the associations between inlining under those optimizations is important for designing the corresponding models.
	
\end{itemize}

Considering the above factors, we decide to train a classifier for ICS prediction. Considering the different inlining decisions under different compilation settings, we treat the ICS prediction as a multi-label classification problem. In MLC, the presence/absence of multiple labels is predicted and multiple labels can be assigned simultaneously to a sample \cite{bogatinovski2022comprehensive}. Usually, there are correlations between these labels. Classifiers, which capture the correlations, can achieve a better performance in prediction.

Thus, to design an MLC model for ICS prediction, we further investigate the correlation between the inlining decisions under different compilations. Based on the findings, we propose an MLC model named ECOCCJ48 (Ensemble of Compiler Optimization based Classifier Chains built with J48) for ICS prediction.

\subsubsection{Correlation between the inlining decisions}

Table \ref{tab:correlation_cross_opt} shows the intersections and differences between the inlining decisions under the Ox optimizations of two compilers, gcc-8.2.0 and clang-7.0. In this table, \textit{opt} is the abbreviation for ``optimization'', \textit{opt1-opt2} means the number of call sites that are inlined when using opt1 but not inlined when using opt2, and \textit{opt1\&opt2} means the number of call sites that are both inlined when using opt1 and opt2. 

\begin{table*}[h]
	\centering
	\caption{Correlation between the inlining decisions cross optimizations}
	\vspace{-5pt}
	\scalebox{1}{
		\begin{tabular}{cc|ccc|ccc}
			\hline
			\multicolumn{2}{c|}{compiler}                   & \multicolumn{3}{c|}{gcc-8.2.0}                                               & \multicolumn{3}{c}{clang-7.0}                                                \\ \hline
			\multicolumn{1}{c|}{opt1}                & opt2 & \multicolumn{1}{c|}{opt1-opt2} & \multicolumn{1}{c|}{opt1\&opt2} & opt2-opt1 & \multicolumn{1}{c|}{opt1-opt2} & \multicolumn{1}{c|}{opt1\&opt2} & opt2-opt1 \\ \hline
			\multicolumn{1}{c|}{\multirow{3}{*}{O0}} & O1   & \multicolumn{1}{c|}{32}        & \multicolumn{1}{c|}{88}         & 5265      & \multicolumn{1}{c|}{16}        & \multicolumn{1}{c|}{83}         & 83        \\ \cline{2-8} 
			\multicolumn{1}{c|}{}                    & O2   & \multicolumn{1}{c|}{45}        & \multicolumn{1}{c|}{75}         & 8919      & \multicolumn{1}{c|}{8}         & \multicolumn{1}{c|}{91}         & 16949     \\ \cline{2-8} 
			\multicolumn{1}{c|}{}                    & O3   & \multicolumn{1}{c|}{34}        & \multicolumn{1}{c|}{86}         & 15180     & \multicolumn{1}{c|}{8}         & \multicolumn{1}{c|}{91}         & 17323     \\ \hline
			\multicolumn{1}{c|}{\multirow{2}{*}{O1}} & O2   & \multicolumn{1}{c|}{352}       & \multicolumn{1}{c|}{5001}       & 3993      & \multicolumn{1}{c|}{38}        & \multicolumn{1}{c|}{128}        & 16912     \\ \cline{2-8} 
			\multicolumn{1}{c|}{}                    & O3   & \multicolumn{1}{c|}{376}       & \multicolumn{1}{c|}{4977}       & 10289     & \multicolumn{1}{c|}{37}        & \multicolumn{1}{c|}{129}        & 17285     \\ \hline
			\multicolumn{1}{c|}{O2}                  & O3   & \multicolumn{1}{c|}{179}       & \multicolumn{1}{c|}{8815}       & 6451      & \multicolumn{1}{c|}{172}       & \multicolumn{1}{c|}{16868}      & 546       \\ \hline
	\end{tabular}}
	\label{tab:correlation_cross_opt}
\end{table*}

When comparing the inlining decisions in lower optimizations with the higher optimizations, we find that a large proportion of inlining decisions in lower optimization also appear when using higher optimizations. For example, when opt1 is O1 and opt2 is O2, 5001 of the inlining decisions conducted in O1 are also conducted in O2, while only 352 of them are not. In general, this proportion is 94.78\% (19042 in 20090), indicating an incremental trend of inlining decisions when increasing the optimizations. 

\begin{figure}[t]
	\centering
	\vspace{-5pt}
	\includegraphics[width=0.55\textwidth]{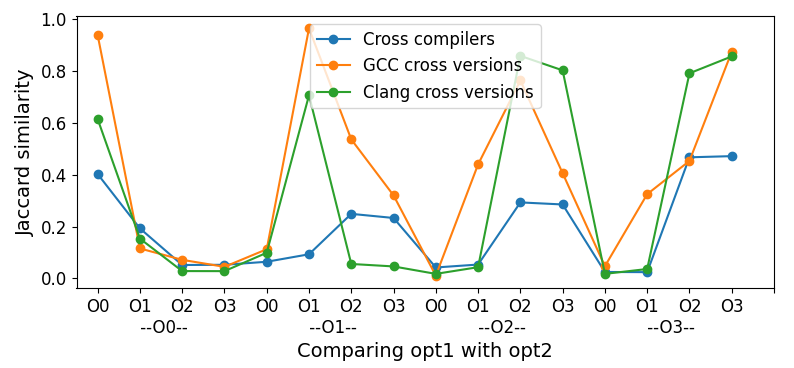}
	\vspace{-8pt}
	\caption{Inlining correlation across compilers and compiler versions}
	\label{fig:correlation_cross_compiler}
	\vspace{-10pt}
\end{figure}

After analyzing the correlations between optimizations in the same compiler, we further compare inlining decisions across compilers and compiler versions. Figure \ref{fig:correlation_cross_compiler} shows the Jaccard similarities between the decisions made by different compilers. In detail, we analyze the results of three compiler pairs: GCC compilers in different versions, Clang compiler in different versions, and the GCC compiler with the Clang compiler.

When comparing the inlining decisions of the compilers in the same family, we found that their inlining decisions are similar to each other, which is consistent with the results observed in the \cite{jia2021one}. As shown in Figure \ref{fig:correlation_cross_compiler}, when applying O0-O3 to gcc-6.4.0 and gcc-7.2.0, 87\% (19087 in 21932) of the call sites present the same inlining decisions, which indicates a large degree of similarity for inlining decisions in the same compiler family.

Though a clear correlation is shown in the optimizations of the same compiler and compilers in the same family, the correlation for different compiler families (GCC and Clang) is ambiguous as shown in Figure \ref{fig:correlation_cross_compiler}. No correlation between GCC and Clang is presented, as GCC and Clang have different considerations for inlining and apply different inlining strategies.

\subsubsection{ECOCCJ48 as the MLC model} Considering the independence of inlining decisions across different compiler families, similar inlining decisions in the same compiler family, and incremental inlining decisions across the optimizations, we design an MLC classifier named ECOCCJ48, which uses binary relevance to predict labels for different compiler families and use the ensemble of order-determined classifier chains to predict labels for different optimizations under the same compiler.

\begin{figure*}[ht]
	\centering
	\vspace{-10pt}
	\subfigure[Compiler-level classifier]{
		\begin{minipage}[t]{0.45\linewidth}
			\centering
			\includegraphics[width=0.64\textwidth]{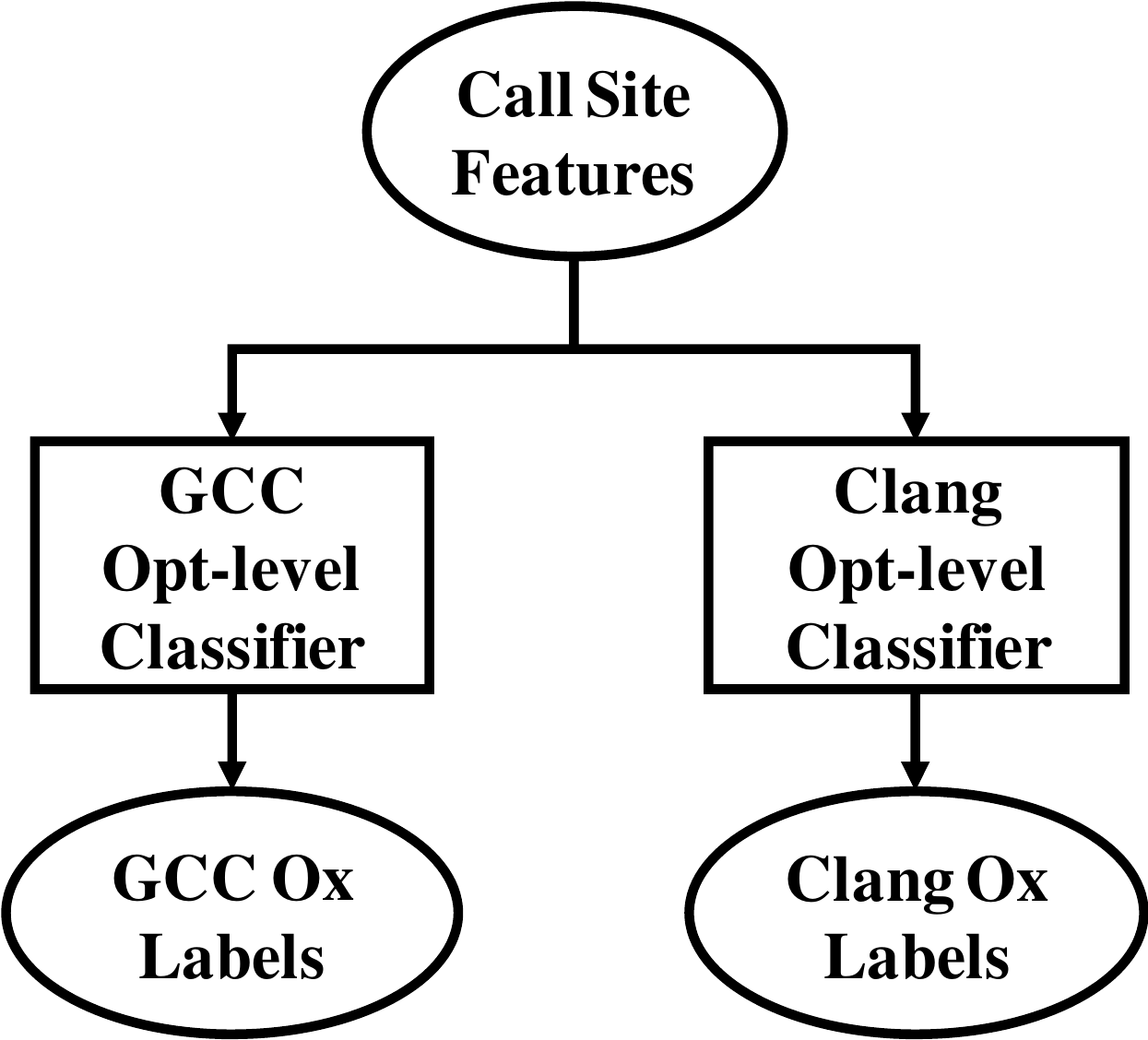}
			\label{fig:ecoccj48_compiler}
			\vspace{-13pt}
		\end{minipage}%
	}%
	\subfigure[Opt-level classifier (in Clang or GCC)]{
		\begin{minipage}[t]{0.45\linewidth}
			\centering
			\includegraphics[width=0.7\textwidth]{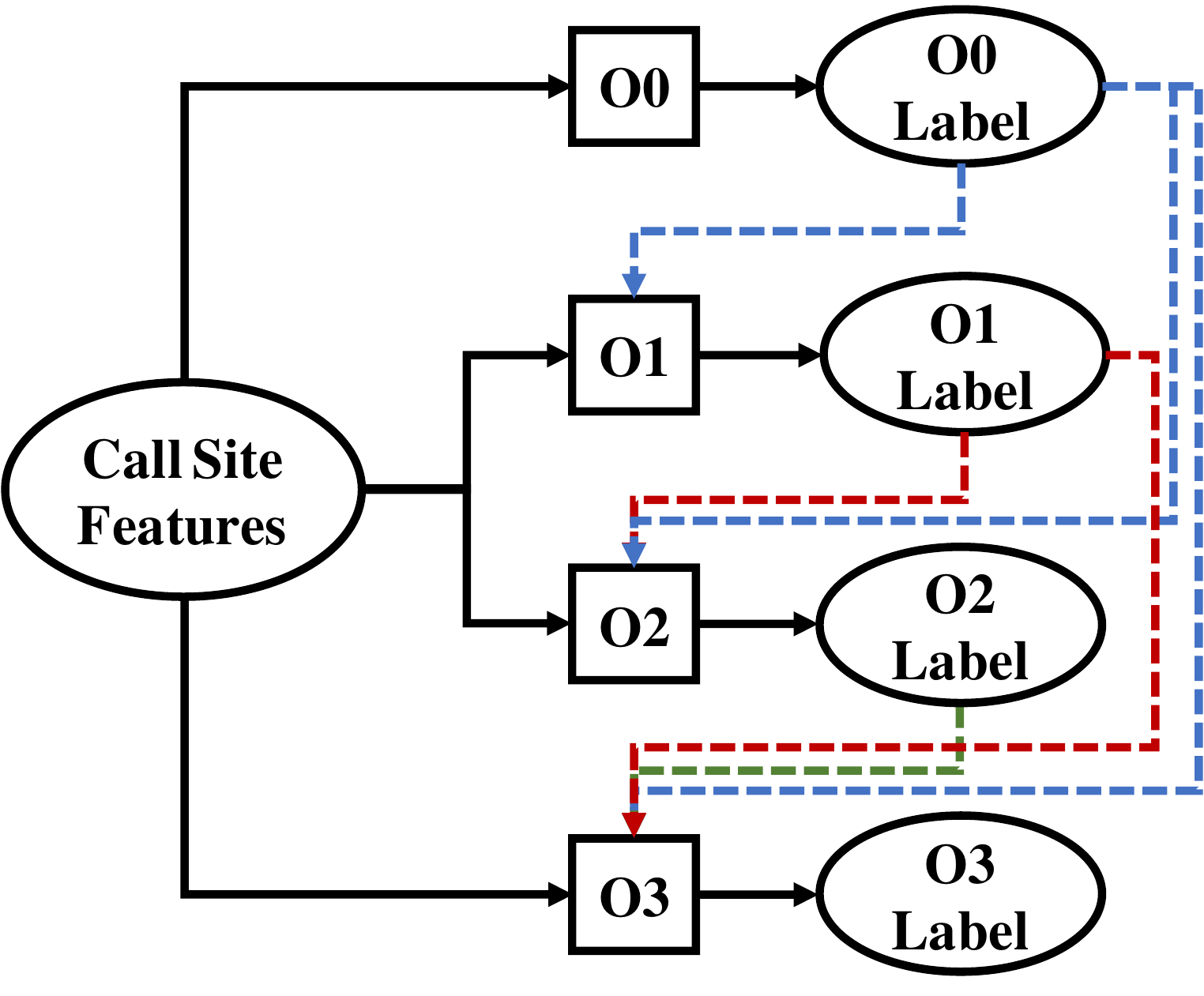}
			\label{fig:ecoccj48_opt}
			\vspace{-13pt}
		\end{minipage}%
	}%
	\vspace{-13pt}
	\caption{Architecture of the classifier model ECOCCJ48}
	\label{fig:ecoccj48}
	\vspace{-15pt}
\end{figure*}

As shown in Figure \ref{fig:ecoccj48}, ECOCCJ48 has a compiler-typical classifier for each compiler family (GCC and Clang) and the labels under these two compilers are not used for each other. Instead, in each compiler-typical classifier (opt-level classifier), labels have a sequential dependence across the optimizations. For example, in Figure \ref{fig:ecoccj48_opt}, the prediction for O2 is based on call site features and labels that have been predicted for O0 and O1. Considering that inlining decisions conducted in O0 and O1 usually appear in O2, the architecture of ECOCCJ48 can leverage the correlations between labels to produce a more accurate prediction result.

\subsubsection{Training on imbalanced dataset}
Though we have developed an MLC model to capture the internal inlining patterns, training it on the labeled dataset still faces two challenges. On the one hand, the majority of call sites are not inlined during compilation and only about 20\% of the call sites will be inlined when using Ox optimizations. On the other hand, the proportion of inlined call sites also varies when compiling using different optimizations. This caused the dataset imbalanced and the training biased toward normal call sites. 

We use ensemble methods to handle the imbalanced dataset. The ensemble method trains the base classifiers on the randomly selected dataset and predicts the labels by aggregating the predictions from the base classifiers. As different base classifiers are trained in different corpora, they can capture the inlining patterns of some rare labels. Besides, ensemble learning methods are superior in terms of performance to single-model learning approaches \cite{moyano2018review}. Though the computational time ensemble learning methods require for building an ensemble is larger than the time for building a single model, our base classifier J48 \cite{salzberg1994c4} has a low computation complexity so that the cost is acceptable for our inlining prediction task.

The architecture of ECOCCJ48 also helps tackle the imbalance under different compilations. For example, the prediction for O3 is based on the predictions of O0, O1, and O2. Leveraging these predictions, the classifier under O3 can also accurately learn the inlining patterns that are already learned in O0, O1, and O2 but may be rare which is difficult to be learned in O3. Thus, ECOCCJ48 can better tackle these rarely inlined cases and produce more complete results.

\section{SFS generation process of O2NMatcher}
\label{sec:SFS_generation}

In Section \ref{sec:classifier}, we have trained a MLC model for ICS prediction. In this Section, we will illustrate the process of generating SFSs for uncompilable OSS projects. 

\subsection{FCG Construction and ICS Prediction}
Given an uncompilable OSS project, we first use a parser to extract all call sites and construct its FCG using the functions as the nodes and calls as the edges. As we consider the locations of call site as an important feature for ICS prediction, the FCG is constructed as a multi-digraph which have multiple directed edges between two nodes and the edges carry the location information of the call sites. 

Then for each call site in the FCG, we extract the features from its caller, callee, and call instruction to form its call site features. Using these features as input to ECOCCJ48, we obtain the labels of each call site under all compilation settings. Considering each compilation separately, we obtain the same number of the labeled FCGs as the number of compilation settings.

\subsection{SFS Generation}
\label{sec:SFS_generation_subsection}

\begin{figure}[ht]
	\centering
	\vspace{-5pt}
	\subfigure[Labeled FCG]{
		\begin{minipage}[t]{0.25\linewidth}
			\centering
			\includegraphics[width=0.7\textwidth]{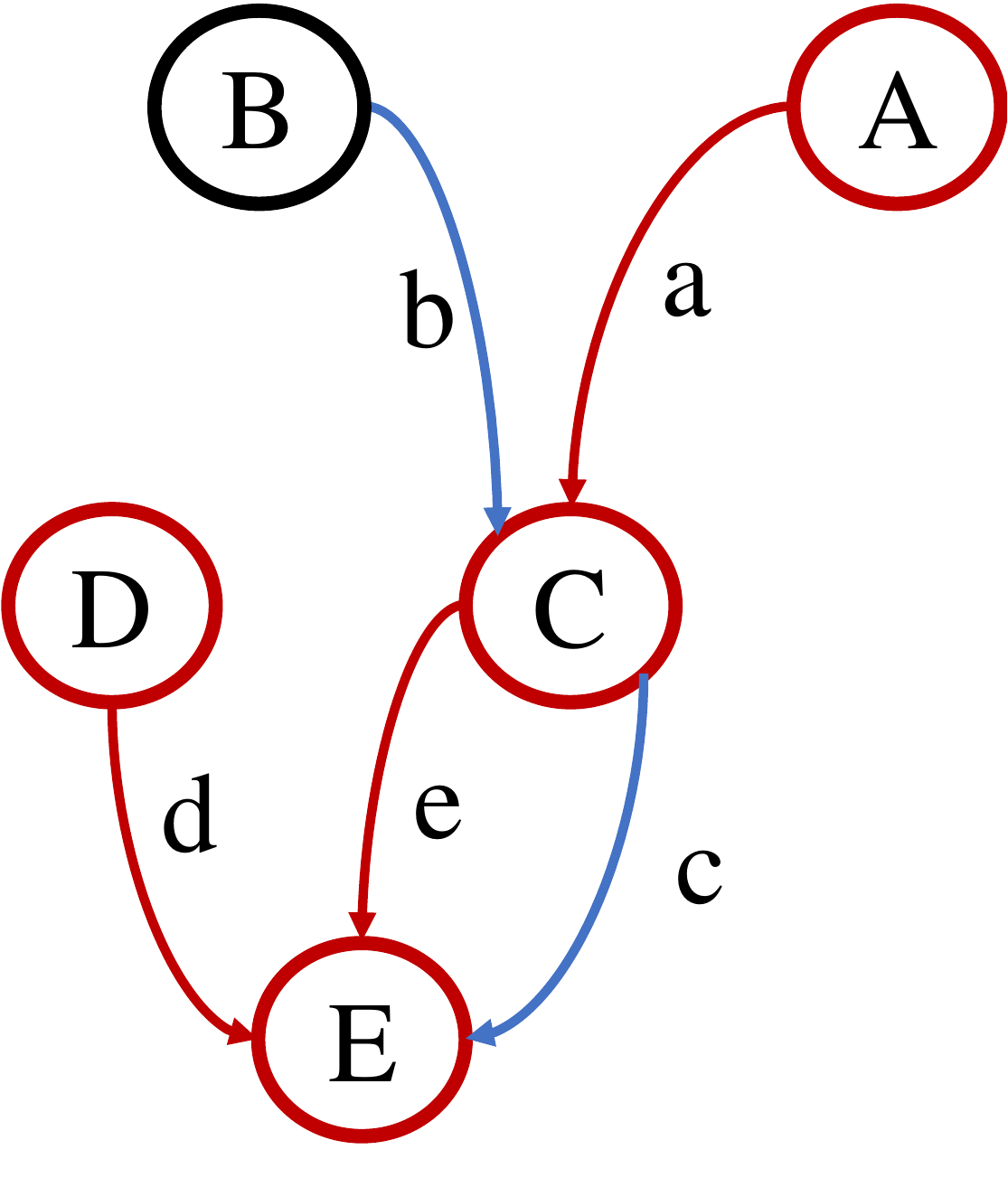}
			\label{fig:SFS_1}
			\vspace{-18pt}
		\end{minipage}%
	}%
	\centering
	\subfigure[Selected root nodes]{
		\begin{minipage}[t]{0.25\linewidth}
			\centering
			\includegraphics[width=0.7\textwidth]{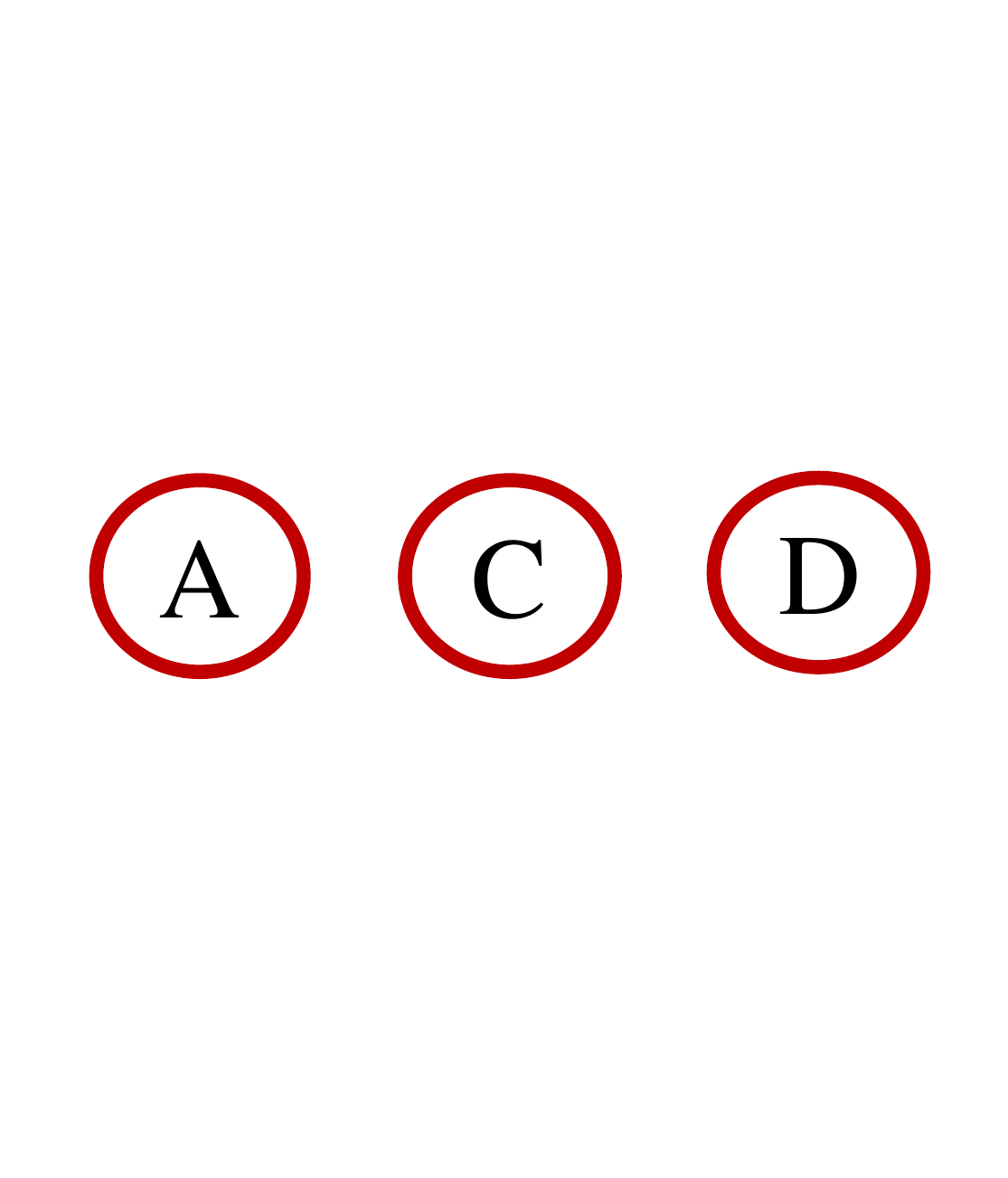}
			\label{fig:SFS_2}
			\vspace{-18pt}
		\end{minipage}%
	}%
	\subfigure[Generated SFSs]{
		\begin{minipage}[t]{0.35\linewidth}
			\centering
			\includegraphics[width=0.85\textwidth]{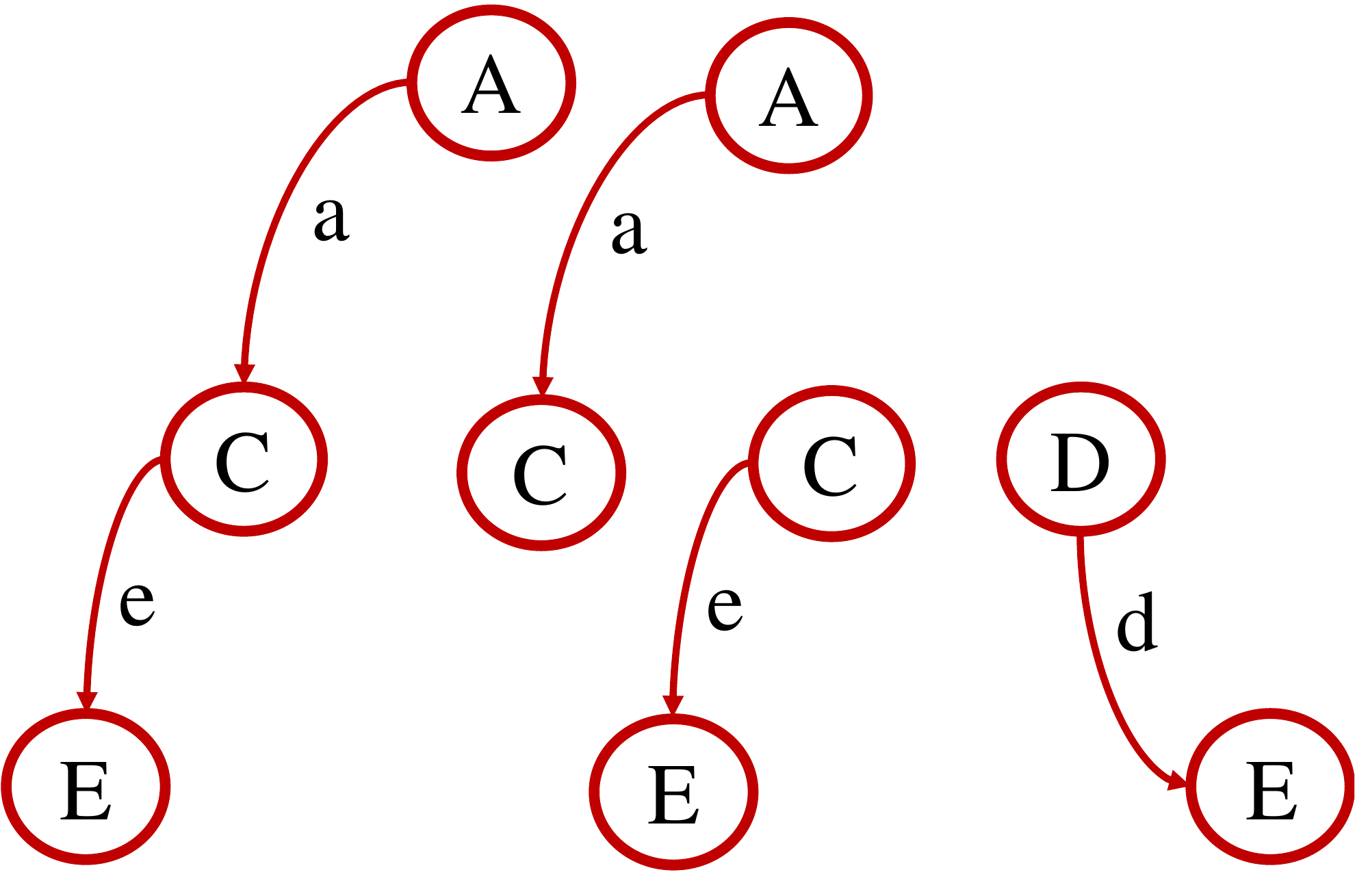}
			\label{fig:SFS_3}
			\vspace{-18pt}
		\end{minipage}%
	}%
	\vspace{-8pt}
	\caption{Example for SFS generation based on labeled FCG}
	\label{fig:SFS_construction_example}
	\vspace{-12pt}
\end{figure}

After we obtained the labeled FCGs, we need to generate the corresponding SFSs in it. Figure \ref{fig:SFS_1} shows an example of labeled FCGs. We use the red edges to represent the inlined calls and the blue edges to represent the normal calls. The red circles represent the functions with inlined calls and the black circles represent functions without. 

Assuming that the labeled FCG is a directed acyclic graph, SFS generation is to select the inlining sub-trees from the labeled FCGs. Thus, we split our SFS generation process into two stages: root node selection and edge extension.

\subsubsection{Root node selection} Firstly, we define the inlining sub-graph as the sub-graph of labeled FCG which is constructed by the inlined calls and their connected functions. Simply, as long as the node can satisfy one of the following two requirements, it can serve as the root node to generate a sub-tree. 

\begin{itemize}
	\item [(1)]
	The node is the root node of the inlining sub-graph.  
	
	\item [(2)]
	The node is a non-root node of the inlining sub-graph. It has inlined calls to other nodes and there are other nodes that have normal calls to it.  
	
\end{itemize}

If a node does not satisfy these two requirements, it is either a node that has no inlined edges to other nodes or a node in the inlining sub-graph but is only called by the inlined edges. The former has no nodes that can be inlined to the SFS while the latter will always be inlined to its caller.

Figure \ref{fig:SFS_1} presents a labeled FCG ready for SFS generation.
Using these two requirements, we can easily select three nodes: A, C, and D, as shown in Figure \ref{fig:SFS_2}. 

\subsubsection{Edge extension} There are also two rules for edge extension.

\begin{itemize}
	\item [(1)]
	If there are only inlined edges between all the callers and callees, traverse the inlining sub-graph and include all the nodes that the root node can reach. 
	
	\item [(2)]
	If there are two different kinds of edges between a caller and a callee, two versions of SFSs will be generated. One follows the inlined edge to include subsequent nodes while one stops at the normal edge.
	
\end{itemize}

When using D as the root node, there are only inlined edges in the sub-tree. Thus, SFS \textit{D$\rightarrow$E} is produced for root node D. Instead, there are two edges c and e between nodes C and E. Thus, we generate the SFS \textit{A$\rightarrow$C$\rightarrow$E} and \textit{A$\rightarrow$C} for root node A and the SFS \textit{C$\rightarrow$E} for root node C, as shown in the Figure \ref{fig:SFS_3}.

If the labeled FCG is a directed cyclic graph, we only need to add a new rule to the edge extension that when traversing the inlining sub-graph, if a callee node is already included in the SFS, this node should be abandoned to avoid duplication in the generated SFSs.

\subsection{SFS Aggregation}
After the generation of SFSs, we need to aggregate the functions in the SFS to be compared with the binary functions. Directly, there are two ways to aggregate the SFS: conduct inlining for the source functions in the SFSs such as the inlining during compilation or directly aggregate the content of all the functions in the SFS.

Though the former way seems to be a better solution to generate a more similar aggregated SFS with the binary function, several inspections motivate us to select the latter. 

Firstly, if a caller function has two inlined call sites to a callee function, it is difficult to decide to inline it at which call site or at both call sites. That is because, when the compiler decides to inline these two call sites, it may consider adjusting the control flow in the caller and merging these two call sites. Thus, only one copy of the callee function is introduced to reduce the cost of inlining functions. 

Secondly, if the labeled FCG is a cyclic graph, where a callee is predicted to be inlined into the caller and the caller also should be inlined to the callee, it is difficult to arrange the inlining sequence and result of these two functions. Instead, just putting these functions together does not face this problem and it also maintains the semantics of the caller and callee functions.

Thirdly, existing binary2source function matching works usually leverage embeddings to calculate the function similarities. We found that the existing works are strong enough to learn and aggregate semantics from two separate source functions. In our experiments, the two ways to aggregate SFS achieve similar performance. Thus simply aggregating the functions saves additional effort and raises a low cost for SFS aggregation. 

Considering the above reasons, we decide to simply aggregate the content of the functions in the SFS to generate the final SFS for comparison.

\section{Evaluation}
\label{sec:evaluation}
In this section, we will first introduce the setup of our experiments. Then, we answer the following research questions to validate the performance of O2NMatcher and ECOCCJ48.

\newtheorem{question}{\textbf{RQ}}
\newcounter{myrq}
\setcounter{myrq}{1}
\renewcommand\themyrq{\arabic{myrq}}
\renewcommand\thequestion{\themyrq}

\begin{question}
	\label{RQ-1}
	\textit{Can O2NMatcher promote the performance of existing 1-to-1 binary2source matching works?}
	
\end{question}
\addtocounter{myrq}{1}
\begin{question}
	\label{RQ-2}
	\textit{How accurate are the SFSs generated by O2NMatcher, compared to Bingo and ASM2Vec?}
\end{question}
\addtocounter{myrq}{1}
\begin{question} 
	\label{RQ-3}
	\textit{How good is the performance of ECOCCJ48, compared to existing multi-label classification works?}
	
\end{question}
\addtocounter{myrq}{1}
\begin{question}
	\label{RQ-4}
	\textit{How much time does O2NMatcher cost in training, predicting, and generating SFSs?}
	
\end{question}
\addtocounter{myrq}{1}
\begin{question}
	\label{RQ-5}
	\textit{What is the contribution of each feature set selected for ICS prediction?}
	
\end{question}
\addtocounter{myrq}{1}

\subsection{Study Setup}

\subsubsection{Evaluation Dataset} To preform a thorough evaluation, we select 51 packages from GNU projects \cite{gnulib}, compiled by 9 compilers, 4 optimizations to X86-64 using Binkit \cite{Binkit}, resulting in 8,460 binaries and 4,294,478 binary functions. The selected projects and versions are shown in Table \ref{tab:dataset} and the select compilers and optimizations are shown in Table \ref{tab:datasetc_compiler_setting}.
This dataset contains many widely-used packages such as \textit{Coreutils} \cite{coreutils} and \textit{OpenSSL} \cite{openssl} which have been extensively used in binary similarity detection works \cite{xu2021interpretation, bingo, liu2018alphadiff, egele2014blanket, wang2017memory, codecmr}.

\begin{table}[h]
	\caption{Projects and Versions in the Dataset}
	\vspace{-8pt}
	\centering
	\begin{tabular}{c|c|c|c|c|c}
		\hline
		Project   & Version & Project       & Version & Project    & Version \\ \hline
		a2ps      & 4.14    & gmp           & 6.1.2   & macchanger & 1.6.0   \\ \hline
		binutils  & 2.3     & gnu-pw-mgr    & 2.3.1   & nettle     & 3.4.1   \\ \hline
		bool      & 0.2.2   & gnudos        & 1.11.4  & osip       & 5.0.0   \\ \hline
		ccd2cue   & 0.5     & grep          & 3.1     & patch      & 2.7.6   \\ \hline
		cflow     & 1.5     & gsasl         & 1.8.0   & plotutils  & 2.6     \\ \hline
		coreutils & 8.29    & gsl           & 2.5     & readline   & 7       \\ \hline
		cpio      & 2.12    & gss           & 1.0.3   & recutils   & 1.7     \\ \hline
		cppi      & 1.18    & gzip          & 1.9     & sed        & 4.5     \\ \hline
		dap       & 3.1     & hello         & 2.1     & sharutils  & 4.15.2  \\ \hline
		datamash  & 1.3     & inetutils     & 1.9.4   & spell      & 1.1     \\ \hline
		direvent  & 5.1     & libiconv      & 1.15    & tar        & 1.3     \\ \hline
		enscript  & 1.6.6   & libidn        & 2.0.5   & texinfo    & 6.5     \\ \hline
		findutils & 4.6.0   & libmicrohttpd & 0.9.59  & time       & 1.9     \\ \hline
		gawk      & 4.2.1   & libtasn1      & 4.13    & units      & 2.16    \\ \hline
		gcal      & 4.1     & libtool       & 2.4.6   & wdiff      & 1.2.2   \\ \hline
		gdbm      & 1.15    & libunistring  & 0.9.10  & which      & 2.21    \\ \hline
		glpk      & 4.65    & lightning     & 2.1.2   & xorriso    & 1.4.8   \\ \hline
	\end{tabular}
	\label{tab:dataset}
\end{table}

\begin{table}[htbp]
	\vspace{-5pt}
	\caption{Compilers and optimizations used in the dataset}
	\label{tab:datasetc_compiler_setting}
	\centering
	\vspace{-8pt}
	\begin{tabular}{c|c}
		\hline
		Compilers     & \begin{tabular}[c]{@{}c@{}}gcc-4.9.4, gcc-5.5.0, gcc-6.4.0, gcc-7.3.0, gcc-8.2.0, \\ clang-4.0, clang-5.0, clang-6.0, clang-7.0\end{tabular} \\ \hline
		Optimizations & O0, O1, O2, O3                                                                                                                               \\ \hline
	\end{tabular}
	\vspace{-5pt}
\end{table}

\subsubsection{Evaluation Benchmark} As our O2NMatcher severs as a complement for existing binary2source function matching works, we need to select one work as the baseline to evaluate our promotion for it. Till now, there are many binary2source matching works \cite{BAT, RESource, JISIS14, BinPro, OSSPolice, Saner2019, B2SFinder, ban2021b2smatcher, codecmr, XLIR}. But among them, CodeCMR achieves the most accurate function-level matching \cite{codecmr}. Besides, CodeCMR has an open-accessed tool named BinaryAI \cite{binaryai}, which facilitates us to carry out further evaluations.

\subsubsection{Experiment Setting.} To evaluate the effectiveness of O2NMatcher, we randomly select 90\% of projects in dataset I as the training dataset and the other 10\% of projects as the testing dataset. As we discovered that inlining decisions in the same compiler tend to be similar, the training dataset is generated only using two compilers: gcc-8.2.0 and clang-7.0, while our testing dataset contains all those 9 compilers. The classifier is trained on the training dataset and produces the SFSs for the testing dataset. To avoid bias, we iterate this process 10 times. 

When running O2NMatcher on the testing dataset, binaries are stripped and the OSS projects cannot be compiled to satisfy the requirement of SCA. We also use the released model of BinaryAI that is merely trained on the normal dataset and evaluate O2NMatcher without retraining it on the inlining dataset.

\subsubsection{Evaluation Metrics} To evaluate the effectiveness of O2NMatcher, we use the recall@K which is widely used in existing binary2source matching works. Recall@k means the proportion of its original function found in the top-k returned source functions. Here, in the ``1-to-1'' matching works, the original function means the source function that the binary function is compiled from. In our O2NMatcher, the SFSs, whose root functions are the original functions, are also considered as a correct return. In the following experiments, we select K as 1, 10, and 50 to evaluate the effectiveness in binary2source function matching tasks.

To evaluate the cost of O2NMatcher, we propose a metric named SFS size to calculate the ratio of the number of generated SFSs to the number of original function sets. The added SFSs increase the size of the corpus, thus it may affect the time of querying and the performance of binary2source matching. 

To evaluate the effectiveness of ECOCCJ48, we select three metrics used in the MLC evaluation, including precision, recall, and F1-score. 
Precision, recall, and F1-score are defined similarly to the single-label classification and are calculated by the weighted values of the metrics for all the labels. 

\subsubsection{Implementation of O2NMatcher} In the dataset labeling, we use Understand to parse source projects and IDA Pro to disassemble binaries. 
In the FCG construction, we use Understand to extract call sites and construct FCGs for source projects and use IDA Pro to construct FCGs for binaries.
In the call site feature extraction, we use tree-sitter to extract function body and function definition features from caller/callee and call instruction features, and use Understand to extract function call features.
In the model training, we use the \textit{scikit-multilearn}\footnote{\url{http://scikit.ml/}}, which is a toolkit \cite{2017arXiv170201460S} in Python, to implement ECOCCJ48 and other MLC methods, using the default parameters. The whole procedure is implemented in Python and we run all the experiments on a workstation equipped with Ubuntu 18.04, Intel Xeon Gold 6266C (44 virtual cores @3GHz), 1024GB DDR4 RAM, and one Nvidia RTX3090 GPU (19500MHz, 24GB GDDR6X, 35.38 TFLOPS FP32).

\subsection{Answer to RQ~\ref{RQ-1}: Effectiveness of O2NMatcher}

In this evaluation, we first compare O2NMatcher with BinaryAI to present the promotion of O2NMatcher on existing ``1-to-1'' matching works. Then we compare O2NMatcher with existing inlining-simulation works to show the effectiveness of O2NMatcher compared with the state-of-the-art works.

\begin{table}[h]
	\caption{Effectiveness of O2NMatcher compared with the state-of-the-art works}
	\vspace{-5pt}
	\begin{tabular}{c|c|c|c|c|c}
		\hline
		Test Set                           & Metrics        & BinaryAI & O2NMatcher      & Bingo            & ASM2Vec          \\ \hline
		\multirow{4}{*}{Inlined Functions} & Recall@1       & 34.50\%  & 40.93\% \textbf{(+6.43\%)} & 34.84\% \textbf{(+0.34\%)} & 34.65\% \textbf{(+0.16\%)} \\ \cline{2-6} 
		& Recall@10      & 51.88\%  & 56.20\% \textbf{(+4.32\%)} & 51.20\% \textbf{(-0.68\%)} & 50.93\% \textbf{(-0.95\%)} \\ \cline{2-6} 
		& Recall@50      & 65.23\%  & 65.75\% \textbf{(+0.52\%)} & 63.63\% \textbf{(-1.60\%)} & 63.66\% \textbf{(-1.57\%)} \\ \cline{2-6} 
		& SFS Size & 0.00\%   & 87.63\%          & 23.78\%          & 22.09\%          \\ \hline
		\multirow{4}{*}{All Functions}     & Recall@1       & 72.84\%  & 73.76\% \textbf{(+0.92\%)} & 72.92\% \textbf{(+0.08\%)} & 72.90\% \textbf{(+0.06\%)} \\ \cline{2-6} 
		& Recall@10      & 87.97\%  & 88.65\% \textbf{(+0.68\%)} & 87.73\% \textbf{(-0.25\%)} & 87.70\% \textbf{(-0.27\%)} \\ \cline{2-6} 
		& Recall@50      & 92.64\%  & 93.03\% \textbf{(+0.39\%)} & 92.41\% \textbf{(-0.22\%)} & 92.40\% \textbf{(-0.24\%)} \\ \cline{2-6} 
		& SFS Size & 0.00\%   & 53.14\%          & 20.77\%          & 18.91\%          \\ \hline
	\end{tabular}
	\vspace{-5pt}
	\label{tab:effectiveness_of_O2NMatcher}
\end{table}

\subsubsection{Compared with BinaryAI.} Table \ref{tab:effectiveness_of_O2NMatcher} shows the effectiveness of O2NMatcher compared with the state-of-the-art works. Firstly, when comparing the performance of O2NMatcher with BinaryAI, we found that O2NMatcher brings a considerable promotion to BinaryAI when detecting inlined functions, where Recall@1 increases by 6.43\%, Recall@10 increases by 4.32\%, and Recall@50 increases by 0.52\%. Though O2NMatcher has to generate 53.14\% more SFSs for comparison, we noticed that O2NMatcher does not degrade the performance of BinaryAI on all the functions. Instead, it increases its Recall@1 by 0.92\%.
Besides, we noticed the SFS size generated for the inlined functions is larger than that for all the functions, which indicates our O2NMatcher can distinguish between the inlined call sites and normal call sites.

\subsubsection{Compared with Bingo and ASM2Vec.} Bingo \cite{bingo} and ASM2Vec \cite{asm2vec} are the two state-of-the-art function-level binary2binary matching methods that target tackling the challenges that function inlining brings. Compared with O2NMatcher which trains the classifier on the labeled datasets, Bingo and ASM2Vec use manual design rules to conduct inlining. They were first designed to resolve the binary2binary matching under function inlining, and we migrate their rules to generate SFSs for binary2source matching. 

The columns under ``Bingo'' and ``ASM2Vec'' present the performance of Bingo and ASM2Vec in promoting existing binary2source matching works on inlined functions. Bingo and ASM2Vec respectively promote BinaryAI by 0.34\% and 0.16\% when detecting the inlined functions. However, the promotion on all the functions is negligible, which is less than 0.1\% for both of them. Besides, this promotion seems to bring a side effect when increasing K in Recall. The performance decreases in Recall@10 and Recall@50 where their generated SFSs decrease the accuracy of normal function matching.

\begin{tcolorbox}
	
	\textbf{Answering RQ~\ref{RQ-1}:}  
	O2NMatcher can promote the performance of existing ``1-to-1'' binary2source matching works by 6\% for functions with inlining. The classifier can differentiate the inlined and normal call sites so that more SFSs are generated towards the functions with inlining. Compared with Bingo and ASM2Vec, O2NMatcher is more effective in helping existing works to promote their performance.
\end{tcolorbox}

\subsection{Answer to RQ~\ref{RQ-2}: Evaluation of generated SFSs}

The fundamental and essential functionality of O2NMatcher is to generate SFSs, and the quality of generated SFSs directly influences the effectiveness of O2NMatcher. In this Evaluation, we compare the SFSs generated by O2NMatcher, Bingo, and ASM2Vec with the ground truth, to reveal their capability of generating SFSs. 

\begin{table}[h]
	\caption{Evaluation of generated SFSs compared with the state-of-the-art works}
	\vspace{-5pt}
	\begin{tabular}{c|c|c|c|c}
		\hline
		Method                          & \multicolumn{1}{l|}{Precision@Func} & \multicolumn{1}{l|}{Recall@Func} & \multicolumn{1}{l|}{Precision@SFS} & \multicolumn{1}{l}{Recall@SFS} \\ \hline
		\multicolumn{1}{l|}{O2NMatcher} & 64.91\%                             & 56.68\%                          & 56.42\%                            & 54.12\%                        \\ \hline
		Bingo                           & 31.55\%                             & 33.94\%                          & 30.51\%                            & 17.07\%                        \\ \hline
		ASM2Vec                         & 32.90\%                             & 30.67\%                          & 27.98\%                            & 15.41\%                        \\ \hline
	\end{tabular}
	\label{tab:sfs}
	\vspace{-5pt}
\end{table}

Table \ref{tab:sfs} presents the precision and recall of identifying source functions that need inlining (Func) and generating SFSs for those functions (SFS). SFSs are generated considering the predicted inlining labels of call sites. As the classifier cannot predict every function accurately, there may be SFSs generated for source functions that do not need them and there also may be source functions that need SFSs but the classier failed to predict. Intuitively, ``Precision@Func'' and ``Recall@Func'' present the precision and recall of identifying source functions that need inlining. When we focus on the source functions that need SFSs and O2NMatcher also generate SFSs for them, ``Precision@SFS'' and ``Recall@SFS'' present the precision and recall of SFSs generated for the source functions that need SFSs. 

As shown in Table \ref{tab:sfs}, when identifying the source functions that need SFSs, we noticed that the precision of O2NMatcher is 32-33\% higher than Bingo and ASM2Vec and the recall is 23-26\% higher than them. When comparing the generated SFSs and the inlined functions in the ground truth, we found that the 56\% of SFSs generated by O2NMatcher have an exact match in the ground truth and O2NMatcher can recover 54\% inlined functions, which are respectively 26-28\% higher and 39-37\% higher than Bingo and ASM2Vec.

Besides the validity of exact matched SFSs, we found that SFSs, which can cover the most of inlined functions for a source function, are also helpful for matching the binary functions with inlining. Here, we further summarize the similarity between these unmatched SFSs and inlined functions in the ground truth, to reveal the quality of these unmatched SFSs.

\begin{figure}[htbp]
	\centering
	\includegraphics[width=0.4\textwidth]{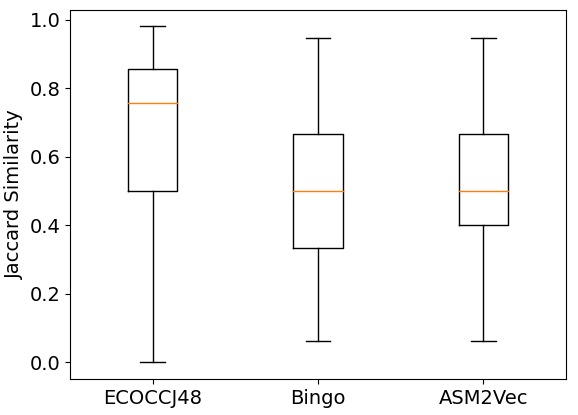}
	\vspace{-5pt}
	\caption{Jaccard similarities between the unmatched SFSs and inlined functions in the ground truth}
	\label{fig:unmatched}
\end{figure}

Figure \ref{fig:unmatched} shows the Jaccard similarity between the unmatched SFSs and inlined functions in the ground truth. Compared with Bingo and ASM2Vec, these unmatched SFSs generated by O2NMatcher are more similar to the sets of inlined functions in the ground truth. In detail, 3/4 of unmatched SFSs generated by O2NMatcher share a similarity of more than 50\% with the inlined functions, and 1/2 of unmatched SFSs share a similarity of more than 78\%, while only 1/4 of SFSs generated by Bingo and ASM2Vec can share a similarity of more than 60\%. Except for those exact matched SFSs, these unmatched SFSs generated by O2NMatcher also share a high similarity with the inlined functions in the ground truth, which also helps promote existing works in detecting these inlined functions.

\begin{tcolorbox}
	\textbf{Answering RQ~\ref{RQ-2}:}  
	Compared with Bingo and ASM2Vec, O2NMatcher can produce more accurate SFSs. On the one hand, O2NMatcher has a 30\% higher precision and 20\% higher recall when identifying the source functions that need SFSs. On the other hand, O2NMatcher has a nearly 30\% higher precision and 40\% higher recall when matching the generated SFSs. Besides, in those unmatched SFSs, SFSs generated by O2NMatcher are also more similar to the inlined functions in the ground truth, which further helps O2NMatcher promote the effectiveness of existing works.
\end{tcolorbox}

\subsection{Answer to RQ~\ref{RQ-3}: Effectiveness of ECOCCJ48}
\label{sec:comparison}

In this section, we compare the ECOCCJ48 with the state-of-the-art multi-label classification methods. Firstly, we will use MLC metrics to evaluate their effectiveness on the ICS prediction. Then we will evaluate their effectiveness in promoting existing ``1-to-1'' binary2source matching works.

Bogatinovski \cite{bogatinovski2022comprehensive} has carried out a comparative study of multi-label classification methods and identifies a subset of five methods that should be used in baseline comparisons: RFPCT (Random Forest of Predictive Clustering Trees) \cite{RFPCT}, RFDTBR (Binary Relevance with Random Forest of Decision Trees) \cite{RFDTBR}, ECCJ48 (Ensemble of Classifier Chains built with J48) \cite{read2010scalable}, EBRJ48 (Ensemble of Binary Relevance built with J48) \cite{read2010scalable}, and AdaBoost \cite{adaboost}. In this evaluation, we will use these five MLC methods as the baseline to compare with ECOCCJ48.

\subsubsection{MLC performance on ICS prediction}

\begin{figure*}[h]
	\centering	
	\subfigure[Precision]{
		\begin{minipage}[t]{0.33\linewidth}
			\centering
			\includegraphics[width=1.8in]{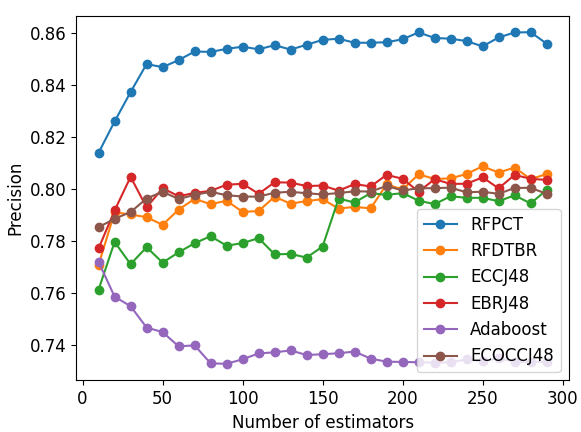}
			\label{fig:precision}
			\vspace{-15pt}
		\end{minipage}%
	}%
	\subfigure[Recall]{
		\begin{minipage}[t]{0.33\linewidth}
			\centering
			\includegraphics[width=1.8in]{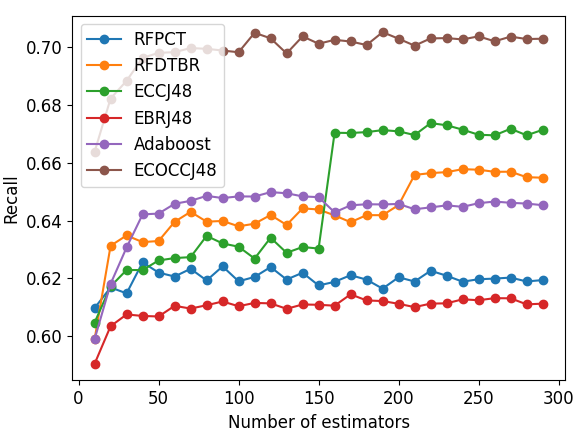}
			\label{fig:recall}
			\vspace{-15pt}
		\end{minipage}%
	}%
	\subfigure[F1-score]{
		\begin{minipage}[t]{0.33\linewidth}
			\centering
			\includegraphics[width=1.8in]{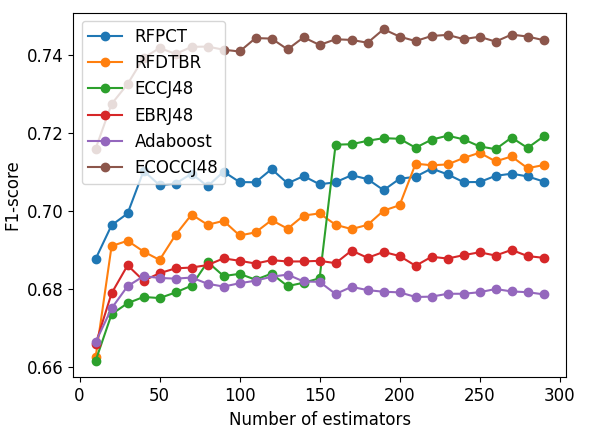}
			\label{fig:f1}
			\vspace{-15pt}
		\end{minipage}%
	}%
	\vspace{-10pt}
	
	\caption{Evaluation results for ECOCCJ48 and its related multi-label classification methods}
	\label{fig:evaluation_for_ECOCCJ48}
	\vspace{-3pt}
\end{figure*}

Figure \ref{fig:evaluation_for_ECOCCJ48} shows the evaluation results on the ICS prediction. The x-axis represents the number of base estimators and the y-axis represents the corresponding metrics. For precision, recall, and F1-score, the higher, the better. For most MLC methods except Adaboost, when increasing the number of estimators, these metrics increase. Adaboost is applying additional estimators to fit for error-classified items, and when the number of estimators increases, Adaboost will be overfitting to the training set and less scalable to the test set.

In summary, when applying the best number of estimators (50 for Adaboost and 300 for others), ECOCCJ48 obtains the best F1-score among all the methods. Though some methods, including RFPCT, RFDTBR, and EBRJ48, obtain a higher precision, ECOCCJ48 obtains a higher recall. Besides, when comparing ECOCCJ48 with RFDTBR, we noticed that ECOCCJ48 can achieve a comparable performance only using a low number of estimators, while RFDTBR achieved that when the number exceeds 200. In summary, ECOCCJ48 can discover more hidden inlined call sites and capture the inlining patterns more quickly.

\subsubsection{MLC performance on binary2source matching}

We also apply these MLC models to the task of binary2source matching by alternating ECOCCJ48 with them. As shown in Table \ref{tab:effectiveness_of_ecoccj48}, ECOCCJ48 achieves three in six times of highest promotion in Recall@1, Recall@10, and Recall@50, while others achieve one or two times of highest promotion. Compared with the state-of-the-art works, ECOCCJ48 can produce more accurate SFSs for binary2source matching, which helps it bring the most considerable promotion when detecting the inlined functions. Besides, other MLC methods can also sever as alternatives to ECOCCJ48 to satisfy different requirements.

\begin{table}[h]
	\caption{Effectiveness of ECOCCJ48 compared with the state-of-the-art works}
	\vspace{-5pt}
	\begin{tabular}{c|c|c|c|c|c|c|c}
		\hline
		Test Set                           & Metrics        & RFPCT            & RFDTBR           & ECCJ48           & EBRJ48  & AdaBoost & ECOCCJ48         \\ \hline
		\multirow{4}{*}{Inlined Functions} & Recall@1       & 40.65\%          & 40.46\%          & 40.77\%          & 40.82\% & 40.50\%  & \textbf{40.93\%} \\ \cline{2-8} 
		& Recall@10      & 55.88\%          & 55.92\%          & \textbf{56.21\%} & 56.20\% & 56.12\%  & 56.20\%          \\ \cline{2-8} 
		& Recall@50      & 65.22\%          & 65.40\%          & 65.66\%          & 65.71\% & 65.38\%  & \textbf{65.75\%} \\ \cline{2-8} 
		& SFS Size & 63.72\%          & 62.73\%          & 82.29\%          & 86.05\% & 76.30\%  & 87.63\%          \\ \hline
		\multirow{4}{*}{All Functions}     & Recall@1       & \textbf{73.77\%} & 73.76\%          & 73.73\%          & 73.74\% & 73.67\%  & 73.76\%          \\ \cline{2-8} 
		& Recall@10      & 88.67\%          & \textbf{88.72\%} & 88.70\%          & 88.65\% & 88.61\%  & 88.65\%          \\ \cline{2-8} 
		& Recall@50      & 93.00\%          & \textbf{93.03\%} & \textbf{93.03\%} & 92.99\% & 92.96\%  & \textbf{93.03\%} \\ \cline{2-8} 
		& SFS Size & 33.25\%          & 32.08\%          & 47.47\%          & 51.42\% & 42.34\%  & 53.14\%          \\ \hline
	\end{tabular}
	\label{tab:effectiveness_of_ecoccj48}
	\vspace{-5pt}
\end{table}

\begin{tcolorbox}
	\textbf{Answering RQ~\ref{RQ-3}:}  
	Compared with the MLC methods, the architecture ECOCCJ48 can discover more hidden inlined call sites and produce more accurate SFSs. As a result, ECOCCJ48 obtained the best F1-score and brought the most considerable promotion on the task of detecting inlined functions.
\end{tcolorbox}

\subsection{Answer to RQ~\ref{RQ-4}: Efficiency of O2NMatcher}

In this section, we further evaluate the efficiency of O2NMatcher by recording the running times at its different stages, including the training time, the predicting time, and the generating time. 

\begin{figure*}[h]
	\centering
	\subfigure[Training]{
		\begin{minipage}[t]{0.33\linewidth}
			\centering
			\includegraphics[width=1.8in]{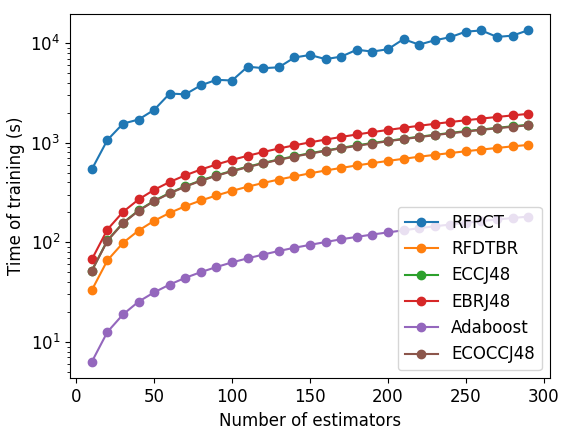}
			\label{fig:time_for_training}
			\vspace{-15pt}
		\end{minipage}%
	}%
	\subfigure[Predicting]{
		\begin{minipage}[t]{0.33\linewidth}
			\centering
			\includegraphics[width=1.8in]{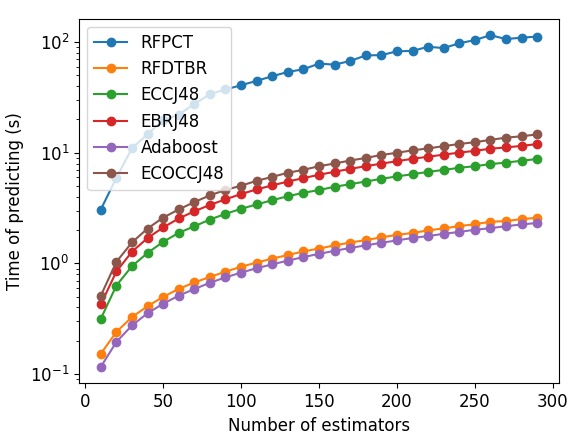}
			\label{fig:time_for_predicting}
			\vspace{-15pt}
		\end{minipage}%
	}%
	\subfigure[Generating]{
		\begin{minipage}[t]{0.33\linewidth}
			\centering
			\includegraphics[width=1.85in]{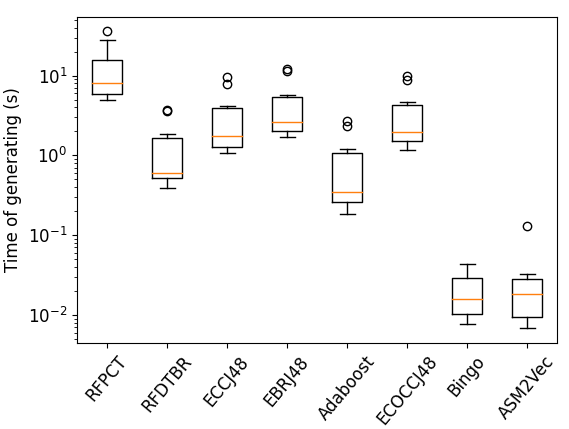}
			\label{fig:time_for_generating}
			\vspace{-15pt}
		\end{minipage}%
	}%
	\vspace{-10pt}
	
	\caption{Running times of O2NMatcher compared with existing works}
	\label{fig:evaluation_time}
	\vspace{-7pt}
\end{figure*}

Figure \ref{fig:time_for_training} shows the training time of ECOCCJ48 and its related MLC methods. The x-axis represents the number of base estimators and the y-axis represents the training time measured in seconds. Among all the MLC methods, it takes the shortest time to train the model for Adaboost with less than 100s while it takes the longest time for RFPCT with more than 10000s. ECOCCJ48 costs a similar time to ECCJ48, where both need 100-1000s to train a model. Considering that model only need to be trained once and can be trained offline, the time cost for training ECOCCJ48 is acceptable.

Figure \ref{fig:time_for_predicting} shows the predicting time of ECOCCJ48 and its related MLC methods. Most of the models only need a few seconds to predict ICS in the test projects. For example, Adaboost trained by 50 estimators only needs 0.4s, and ECOCCJ48 trained by 300 needs 15.2s. Note that ECOCCJ48 can achieve a comparative performance when only using 50 estimators, which will only cost 2.6s for prediction. 

Figure \ref{fig:time_for_generating} shows the SFS generation time for O2NMatcher using different MLC classifiers and two related works, Bingo and ASM2Vec. As Bingo and ASM2Vec determine whether to inline by the features of callees and take the same inlining decisions for all the compilations, they only require 0.01s to generate SFSs. Instead, O2NMatcher must analyze each compilation and traverse the graph to generate SFSs, which raises a higher time cost. Especially, ECOCCJ48 has a higher recall which also brings a larger inlining sub-graph, which subsequently increases the time for the generation. Even so, it only costs less than 10s to generate SFSs for most projects. Besides, the SFS generation progress can be executed in the background, and SFSs are only needed to be generated once before the binary2source matching.

\begin{tcolorbox}
	\textbf{Answering RQ~\ref{RQ-4}:}  
	In general, O2NMatcher can efficiently generate SFSs. Averagely, O2NMatcher spends less than 1000s to train a model, less than 10s to predict ICS, and less than 10s to generate SFSs. Moreover, O2NMatcher is designed to generate SFSs to complement the existing ``1-to-1'' matching. Thus, all these processes only need to be executed once in the background before the binary2source matching.
\end{tcolorbox}

\subsection{Answer to RQ~\ref{RQ-5}: The ablation study on feature selection}

In Section \ref{sec:CS_feature}, we have selected features from three parts of the call site: caller, callee, and call instruction. In this section, we will combine these three features randomly and apply them to the task of ICS prediction. Note that the training projects and testing projects are randomly selected in the ICS prediction, thus the results may have minor changes. But the findings remain the same.

\begin{table}[h]
	\caption{The effectiveness of different feature sets on ICS prediction}
	\vspace{-5pt}
	\begin{tabular}{c|c|c|c}
		\hline
		Feature Set                        & Precision & Recall  & F1-score \\ \hline
		Call Instruction                   & 35.74\%   & 9.00\%  & 13.86\%  \\ \hline
		Caller                             & 46.65\%   & 15.45\% & 22.03\%  \\ \hline
		Caller + Call Instruction          & 46.48\%   & 23.63\% & 29.69\%  \\ \hline
		Callee                             & 68.54\%   & 63.59\% & 65.15\%  \\ \hline
		Callee + Call Instruction          & 69.85\%   & 64.80\% & 66.51\%  \\ \hline
		Caller + Callee                    & 71.73\%   & 66.56\% & 68.48\%  \\ \hline
		Caller + Callee + Call Instruction & 72.20\%   & 66.56\% & 68.52\%  \\ \hline
	\end{tabular}
	\label{tab:ablation}
	\vspace{-5pt}
\end{table}

Table \ref{tab:ablation} shows the effectiveness of different feature sets on ICS prediction. When only using one feature set for prediction, we noticed that using features of callee achieves the best performance, where the F1-score can reach 65.15\%, even higher than combining the caller features and call instruction features. Bingo and ASM2Vec are mainly using the callee features for inlining candidate selection. However, we noticed that the performance based on callee features can still be promoted by combining it with call instruction features and caller features. In detail, combing call instruction features and callee features brings an increase of 1.36\% in the F1-score, while combing caller features and callee features brings an increase of 3.33\%. Finally, combining these three features achieve the best performance, where ECOCCJ48 earns a 68.52\%  F1-score, which is 3.37\% higher than merely using callee features.

\begin{tcolorbox}
	\textbf{Answering RQ~\ref{RQ-5}:}  
	Of the three call site features selected for ICS prediction, callee features are the most important which achieve the best performance. However, combing callee features with caller features and call instruction features can still promote the effectiveness of ECOCCJ48, which brings a 3.37\% promotion in the F1-score.
\end{tcolorbox}

\section{Discussion}
\label{sec:discussion}

In this section, we will discuss some study settings of O2NMatcher and some promising directions.

\subsection{Study setting}

\textbf{Our labeling method.} In this work, we apply a similar method used in Jia \cite{jia2021one} to label the inlined call sites. But as discussed in Jia \cite{jia2021one}, the static parsing tools used in the labeling method, such as Understand and tree-sitter, may only recover part of source entities and the dependence between them. We also found that some entities are not fully covered using our method. To ensure training accuracy, we only select those fully covered functions and their call sites as our training dataset.

\textbf{Parsing test projects.} Though we can select the full recovered functions for training, we cannot discard any functions or call sites in the test projects. To avoid the error that static paring tools bring, we set several rules to process the results they returns. For example, tree-sitter sometimes may mistakenly identify an ``if'' statement as a function, and we set up a filter to refuse source functions with a function name the same as those statements. However, even that, some errors remain and the FCGs that Understand extracted are not fully precise and complete. Thus, our method is affected by these imprecise results and can only produce a degraded performance.

\textbf{Metrics used when searching commonly used source functions.} In binary2source function matching, when matching the queried binary function compiled by commonly used source functions, there may be multiple returned source functions that share the same semantics with each other. To better evaluate the quality of returns, we follow the setting of BinaryAI and calculate duplicate Recall@1 for those commonly used source functions. The duplicate Recall@1 is that if the right return shares the same similarity with the top1 return, it is also counted as a correct return. The setting of duplicate Recall@1 avoids the searching cases where a queried binary function has multiple source functions.

\subsection{Promising directions}

\textbf{Searching for inlined source functions.} In this paper, we propose O2NMatcher to help existing binary2source works find the corresponding SFS of the queried binary function under inlining. However, there are also applications that one needs to search whether a source function is contained in a binary function. This can be accomplished by O2NMatcher, but it is not our focus in this work. When searching inlined source functions in the target binary functions, we regard it as a parts-to-whole matching problem. It is also an important study in source2binary matching and we leave it as future work.

\textbf{Partial inlining.} Clang and GCC have applied partial inlining to inline a part of the callee function into the caller function \cite{gcc_partial, clang_partial}. The partial inlining may bring new challenges as determining when partial inlining happens and which part of the callee function will be inlined are unknown. Our dataset and evaluation have already included the occurrence of partial inlining. For example, compiling projects using gcc in O2 will enable a flag named \textit{-fpartial-inlining}\cite{gcc_options}. However, in the work, we mainly focus on the occurrence of normal inlining and we leave the study of partial inlining as future work.

\section{Related Work}

In this section, we will discuss related works in two areas: binary code similarity analysis under function inlining and multi-label classification.

\subsection{Binary code similarity analysis under function inlining}

Before our work, there are some studies that also investigate the problem associated with function inlining and binary code similarity analysis. Our study is related to them and based on some of their findings.

Bingo and ASM2Vec are two works to resolve binary2binary matching under function inlining. Bingo \cite{bingo} proposes a selective inlining strategy to solve function inlining for binary2binary function matching. ASM2Vec \cite{asm2vec} adds some restrictions to Bingo's strategy for controlling the scope of functions that need inlining. They conduct inlining for binary functions and compare those binary functions after inlining to obtain similarities.

Jia \cite{jia2021one} conduct a study to investigate the effect of function inlining on binary code similarity works. They propose a method to identify function inlining, analyze the existence of function inlining, and point out the effect of function inlining on binary similarity works. Bingo and ASM2Vec are also evaluated and results show they can only recover 60\% of inlined functions. Besides, their inlining decisions are too redundant that only 20\%-30\% of them are necessary.

Different from these works, our work for the first time investigates the binary2source matching under function inlining. Especially, we model the prediction of inlined call sites as a multi-label classification problem, which can produce more accurate SFSs and can help future works better understand the method to tackle function inlining.

\subsection{Multi-label classification}

As mentioned in \cite{zhang2013review}, the multi-label classification methods can be classified into two categories: problem transformation methods and algorithm adaptation methods.

Problem transformation methods tackle the multi-label learning problem by transforming it into other well-established learning scenarios. For example, Binary Relevance \cite{boutell2004learning} and Classifier Chains \cite{read2011classifier} transform the task of multi-label learning into the task of binary classification. Label Powerset transforms \cite{RFDTBR} the MLC method into a multi-class classification problem and Calibrated Label Ranking \cite{furnkranz2008multilabel} transforms the task of multi-label learning into the task of label ranking. Besides, ensemble techniques such as stacking, bagging, and random sub-spacing are also used to promote the performance of single multi-label learning models. As mentioned in \cite{bogatinovski2022comprehensive}, ensemble methods are usually more effective than single methods.

Algorithm adaptation methods tackle the multi-label learning problem by adapting popular learning techniques to deal with multi-label data directly. For example, ML-kNN \cite{zhang2007ml} adapts lazy learning techniques and ML-DT \cite{clare2001knowledge} adapts decision tree techniques. Besides, Rank-SVM \cite{elisseeff2001kernel} adapts kernel techniques and CML \cite{ghamrawi2005collective} adapts information-theoretic techniques.

In Section \ref{sec:comparison}, we select RFPCT, RFDTBR, ECCJ48, EBRJ48 and Adaboost for comparison. These methods show the best predictive performance on average across the different problems \cite{bogatinovski2022comprehensive}. Among them, RFDTBR, ECCJ48, EBRJ48, and Adaboost are problem transformation methods, while RFPCT is an algorithm adaptation method.

In our work, we have carried out several analyses of the correlations of inlining decisions in different compilation settings to design a multi-label classification model ECOCCJ48 for ICS prediction. Compared to the existing multi-label classification methods, ECOCCJ48 takes more consideration of inlining correlations across compilations and thus presents a better performance in predicting inlined call sites.

\section{Conclusion}

In this paper, we proposed a method named O2NMatcher to help conduct binary2source function matching under function inlining. The key idea of O2NMatcher is to generate Source Function Sets (SFSs) as the matching target for binary functions with inlining. We first propose a model named ECOCCJ48 for inlined call site prediction. To train this model, we leverage the compilable OSS to generate a dataset with labeled call sites (inlined or not), extract several features from the call sites, and design a compiler-opt-based multi-label classifier by inspecting the inlining correlations between different compilations. Then, we use this model to predict the labels of call sites in the uncompilable OSS projects without compilation and obtain the labeled function call graphs of these projects. Next, we regard the construction of SFSs as a sub-tree generation problem and design root node selection and edge extension rules to construct SFSs automatically. Finally, these SFSs will be added to the corpus of source functions and compared with binary functions with inlining. We conduct several experiments to evaluate the effectiveness of O2NMatcher and results show our method increases the performance of existing works by 6\% and exceeds all the state-of-the-art 
works. 

We hope that our work can help existing binary2source matching works tackle the challenges that function inlining brings and can provide insights for subsequent research.

\bibliographystyle{ACM-Reference-Format}
\bibliography{references}

\end{document}